\documentclass[longauth]{aa}
\usepackage{graphicx}
\usepackage{txfonts}
\usepackage{amsmath,amssymb}
\usepackage{natbib}

\usepackage[squaren,Gray]{SIunits}
\usepackage{color, colortbl}
\usepackage{array}
\usepackage{hyperref}
\usepackage{longtable,lscape}
\usepackage{multirow}

\newcommand{\Mearth}{$M_{\oplus}$}
\newcommand{\Mjup}{$M_{\text{Jup}}$}

\newcommand\Lp{$\text{L}^\prime$}

\newcommand {\micron}{\unit{}{\micro\meter}}

\begin{document} 

\title{Near-infrared scattered light properties of the HR\,4796\,A dust ring}
\subtitle{A measured scattering phase function from $13.6^\circ$ to $166.6^\circ$}

   \author{
   	J. Milli   \inst{1,2}
	\and A. Vigan \inst{3}
        \and D. Mouillet   \inst{2}
	\and	A.-M. Lagrange \inst{2} 
      	\and J.-C. Augereau \inst{2}
         \and C. Pinte \inst{2,4}
         \and D. Mawet \inst{5,6}
         \and H.~M. Schmid \inst{7}
         \and A. Boccaletti \inst{8}
         \and L. Matr\`{a} \inst{9}
         \and Q. Kral \inst{9}
         \and S. Ertel \inst{10}
         \and G. Chauvin \inst{2}
	\and A. Bazzon \inst{7}
	\and F. M\'enard \inst{2,4}
         \and J.-L. Beuzit \inst{2}
         \and C. Thalmann \inst{7}
         \and C. Dominik \inst{11}
	\and M. Feldt \inst{12}
         \and T. Henning \inst{12}
         \and M. Min \inst{11,13}
         \and J.~H. Girard \inst{1}
         \and R. Galicher \inst{8}
         \and M. Bonnefoy \inst{2}
	\and T. Fusco \inst{14}
	\and J. de Boer \inst{15}
	\and M. Janson \inst{16}
         \and A.-L. Maire \inst{12}
         \and D. Mesa \inst{17}
         \and J. E. Schlieder \inst{12,18}
         \and the SPHERE consortium
          }

   \institute{
   	 European Southern Observatory (ESO), Alonso de C\'ordova 3107, Vitacura, Casilla 19001, Santiago, Chile  \\
         \email{jmilli@eso.org}
     	 \and
	 Univ. Grenoble Alpes, CNRS, IPAG, F-38000 Grenoble, France 
          \and 
	 Aix Marseille Univ, CNRS, LAM, Laboratoire d'Astrophysique de Marseille, Marseille, France
	  \and
	   UMI-FCA, CNRS/INSU France (UMI 3386), and Departamento de Astronomia, Universidad de Chile,
Casilla 36-D Santiago, Chile	
	\and  		
          Department of Astronomy, California Institute of Technology, 1200 E. California Blvd, MC 249-17, Pasadena, CA 91125 USA
         \and 
         Jet Propulsion Laboratory, California Institute of Technology, 4800 Oak Grove Drive, Pasadena, CA 91109, USA
	\and
	ETH Zurich, Institute for Astronomy, 8093 Zurich, Switzerland
	 \and
          LESIA, Observatoire de Paris, PSL Research University, CNRS, Sorbonne Universit\'es, UPMC Univ. Paris 06, Univ. Paris Diderot, Sorbonne Paris Cit\'e
          \and
          Institute of Astronomy, University of Cambridge, Madingley Road, Cambridge CB3 0HA, UK
	\and
	Steward Observatory, University of Arizona, 933 N Cherry Ave, Tucson, AZ 85719, USA
          \and 
          Anton Pannekoek Astronomical Institute, University of Amsterdam, PO Box 94249, 1090 GE Amsterdam, The Netherlands
          \and 
          Max-Planck-Institut f\"ur Astronomie, K\"onigstuhl 17, 69117 Heidelberg, Germany
           \and 
           SRON Netherlands Institute for Space Research, Sorbonnelaan 2, 3584 CA Utrecht, The Netherlands
          \and 
          ONERA, The French Aerospace Lab, BP72, 29 avenue de la Division Leclerc, 92322 Chatillon Cedex, France
          \and
          Leiden Observatory, Leiden University, P.O. Box 9513, 2300 RA Leiden, The Netherlands
          \and 
          Stockholm University, AlbaNova University Center, Department of Astronomy, SE-106 91 Stockholm
         \and
          INAF Osservatorio Astronomico di Padova, Vicolo dell'Osservatorio 5, 35122 Padova, Italy
          \and
          NASA Exoplanet Science Institute, Caltech, Pasadena, California, USA
             }

   \date{Received: 25 November 2015; accepted: 16 December 2016 }

% \abstract{}{}{}{}{} 
% 5 {} token are mandatory
 
  \abstract
   {HR\,4796\,A is surrounded by a debris disc, observed in scattered light as an inclined ring with a high surface brightness. Past observations have raised several questions. First, a strong brightness asymmetry detected in polarised reflected light has recently challenged our understanding of scattering by the dust particles in this system. Secondly, the morphology of the ring strongly suggests the presence of planets, although no planets have been detected to date.}
   {We aim here at measuring with high accuracy the morphology and photometry of the ring in scattered light, in order to derive the phase function of the dust and constrain its near-infrared spectral properties. We also want to constrain the presence of planets and set improved constraints on the origin of the observed ring morphology.}
   {We obtained high-angular resolution coronagraphic images of the circumstellar environment around HR\,4796\,A with VLT/SPHERE during the commissioning of the instrument in May 2014 and during guaranteed-time observations in February 2015. The observations reveal for the first time the entire ring of dust, including the semi-minor axis that was previously hidden either behind the coronagraphic spot or in the speckle noise. }
   {We determine empirically the scattering phase function of the dust in the H band from $13.6^\circ$ to $166.6^\circ$. It shows a prominent peak of forward scattering, never detected before, for scattering angles below $30^\circ$. We analyse the reflectance spectra of the disc from the 0.95\micron{} to 1.6\micron, confirming the red colour of the dust, and derive detection limits on the presence of planetary mass objects.}
  {We confirm which side of the disc is inclined towards the Earth. The analysis of the phase function, especially below $45^\circ$, suggests that the dust population is dominated by particles much larger than the observation wavelength, of about 20\micron. Compact Mie grains of this size are incompatible with the spectral energy distribution of the disc, however the observed rise in scattering efficiency beyond $50^\circ$ points towards aggregates which could reconcile both observables. We do not detect companions orbiting the star, but our high-contrast observations provide the most stringent constraints yet on the presence of planets responsible for the morphology of the dust.}


   \maketitle
%
%________________________________________________________________

\section{Introduction}

The system HR\,4796\,A is a unique laboratory to characterise dust in debris discs. Also known as TWA\,11\,A, this A0V star is part of the TW Hydra kinematic group, with an age recently re-estimated to $10\pm3$ Myr-old \citep{Bell2015} and at a distance of 72.8pc \citep{VanLeeuwen2007}. It harbours one of the debris discs with the highest fractional luminosity, shaped as a thin ring of semi-major axis $\sim$ 77 au inclined by $\sim 76^\circ$. It is bound to the M2 companion HR\,4796\,B orbiting at a projected separation of 560 au, and likely part of a tertiary system with an additional M dwarf at a separation of $\sim13500$ au \citep{Kastner2008}. The surrounding dust was first identified by \citet{Jura1991} from the infrared excess of the star, and resolved by \citet{Koerner1998} and \citet{Jayawardhana1998} at mid-infrared wavelengths from the ground. It was then resolved at near-infrared wavelengths with NICMOS on the Hubble Space Telescope (HST) \citep{Schneider1999} and from the ground, with adaptive optics \citep[AO, ][]{Augereau1999,Thalmann2011,Lagrange2012_HR4796,Wahhaj2014,Rodigas2015,Perrin2015,Milli2015} and at visible wavelengths with HST / STIS \citep{Schneider2009}.

Modelling work by \citet{Augereau1999} indicated that planetesimals larger than one metre undergo a collisional cascade, producing dust particles down to a few microns. Submillimetre observations suggest that the system possesses between 0.25\Mearth{} and a few Earth masses of dust \citep{Greaves2000}. Particles below the blowout size limit of $\sim10\micron$ \citep{Augereau1999} are expected to be ejected from the system by the stellar radiation pressure.
The planetesimals producing the dust in debris discs are a natural outcome of the planet formation process. Although there is, to date, no direct detection of a planetary mass object in this system, striking evidence of one or multiple planets interacting with the disc has been found in earlier observations \citep{Lagrange2012,Wahhaj2014}. The ring has steep edges, which is not expected because collisional evolution would cause a sharp ring to spread out with time. It could be explained by the interaction with gas \citep{Lyra2013} or with one or several planets shaping the inner and outer edges \cite[e.g.][]{Wisdom1980, Lagrange2012}. The ring is also eccentric, suggesting that it is being secularly perturbed by an eccentric planet. 
In addition to these intriguing morphological parameters, the observations also challenge our understanding of light scattering by dust particles. The ansae were seen brighter in the east than in the west in unpolarised optical scatterered light \citep{Schneider2009}, but  recent observations in polarised light showed a dramatic opposite asymmetry near the semi-minor axis: the west side is more than nine times brighter than the east side \citep{Milli2015,Perrin2015}. Many possibilities have been discussed to explain the observations: elongated grains larger than $1 \micron{}$, aggregates made of $1 \micron{}$ elementary particles, a non-axisymmetric dust distribution or a marginally optically thick disc. The lack of detailed knowledge on the optical scattering properties is currently the major obstacle to the analysis of these data \citep{Stark2014,Milli2015}

Scattered light observations produce the highest angular resolution images of circumstellar discs, strongly constraining the architecture of the underlying planetary system. We recorded deep coronagraphic images of HR\,4796\,A during the comissioning and early guaranteed-time observations (GTO) of the Spectro-Polarimetric High-contrast Exoplanet Research instrument \cite[SPHERE,][]{Beuzit2008}. We present first the data, the reduction methods and the contrast obtained (Section \ref{sec_obs}), then we measure the morphology in Section \ref{sec_morphology} and the dust properties in Section \ref{sec_dust_prop} including the scattering phase function and dust spectral reflectance. Finally, we discuss the new constraints on the dust properties in Section \ref{sec_discussion_dust_prop} and speculate on the origins of such a sharp offset ring in Section \ref{sec_origin_offset_sharp} before concluding in Section \ref{sec_conclusions}. 

%__________________________________________________________________

\section{Observations and data reduction}
\label{sec_obs}
\subsection{Observations}

Two sets of near-infrared coronagraphic observations obtained on two epochs are presented here, as shown in Table \ref{tab_log}. Both observations used the pupil-tracking mode of SPHERE, to keep the aberrations as stable as possible and benefit from the field rotation to perform angular differential imaging \cite[ADI,][]{Marois2006}. To reach a high contrast, both sets made use of the coronagraphic combination N\_ALC\_YJH\_S corresponding to an apodizer, a Lyot mask of diameter 185\,mas and an undersized Lyot stop to block the starlight rejected off the mask as well as cover the telescope spiders.

The first data set was recorded during the first commissioning of the SPHERE instrument in April 2014\footnote{The HR\,4796\,A image from the April 2014 data set was part of the SPHERE first light images presented in the ESO press release 1417 \href{http://www.eso.org/public/news/eso1417/}{http://www.eso.org/public/news/eso1417/}.}. We used the IRDIS subsystem \citep{Dohlen2008} in classical imaging \citep{Langlois2010} with the broadband H filter ($\lambda=1.625$\micron, $\Delta\lambda=0.29$\micron).The IRDIS imager splits the incoming light in two channels, and in the case of broadband imaging, those two channels record the exact same information. IRDIS provides a 11\arcsec$\times$11\arcsec{} field of view with a pixel scale of 12.25 mas. The star was observed after meridian passage, during 42 minutes under average to poor atmospheric conditions. Because the conditions degraded during the observations, with a coherence time of only 1 ms at the end of the sequence, only the first 27 min were actually used in the data reduction, corresponding to a parallactic angle variation of $21.5^\circ$ out of a total available of $31.2^\circ$. The deep coronagraphic sequence was followed by a point-spread function (hereafter PSF) measurement out of the coronagraphic mask with the neutral density filter ND\_2, and by an acquisition of sky frames. A sequence of coronagraphic images with four \text{satellite spots} imprinted at a separation of $20\lambda/D$ by applying a periodic modulation to the deformable mirror was also recorded to register the location of the star behind the coronagraphic mask. 

A second set of observations was recorded in February 2015 with a different instrumental setup, known as the IRDIFS mode\footnote{Based on observations made at the Paranal Observatory under ESO programme 095.C-0298(H)}. The SPHERE Integral Field Spectrograph \citep[IFS,][]{Claudi2008} recorded spectral cubes of images from the Y band to the J band, while IRDIS recorded simultaneously images with the dual-band filter H2H3 ($\lambda_{H2}=1.593\micron, \lambda_{H3}=1.667$\micron, $\Delta\lambda_{H2}=0.052\micron, \Delta\lambda_{H3}=0.057$\micron) \citep{Vigan2010}. The conditions were much better and much more stable, with a coherence time above 10ms over the whole sequence. The IFS dataset consists of 21000 spectra covering a total field of view of 1.73\arcsec$\times$1.73\arcsec{} and with a native spaxel size of 12.25 mas. The spectral resolution is $\sim50$. This IRDIFS sequence was followed by an unsaturated PSF measurement out of the coronagraphic mask using the neutral density filter ND\_2.

\begin{table*}
%\begin{minipage}[t]{\columnwidth}
\caption{Log of the two sets of SPHERE observations of HR\,4796\,A.}
\label{tab_log}
\centering
%\begin{center}
\renewcommand{\footnoterule}{}  % to avoid a line before footnotes
\begin{tabular}{p{1.5cm} p{1.9cm} p{2.1cm} p{1.3cm} p{1.cm} p{1.5cm} p{0.7cm} p{2.5cm} p{2cm}}
\hline 
Date & Set-up & DIT\tablefootmark{a} (s) x &    Par.  & Seeing &        Coh. &       Wind   & True & Platescale\tablefootmark{d} \\
  &  &  NDIT x NEXP &   angle\tablefootmark{b} ($^\circ$)  &  (")  &  time\tablefootmark{c} (ms) & (m/s)  & north\tablefootmark{d} ($^\circ$)    & (mas/pixel) \\
\hline
\hline
 2014/05/19 & IRDIS H & 3x15x32 & 8.1;39.3 & 0.8;1.2 & 2.3 ; 1.0 & 10.5 & $-134.87 \pm 0.6$ & $12.238 \pm 0.020$ \\
% 2014/05/19 & IRDIS H & 32x8x16 & 8 / 32 & 1.04/1.06 & 1.2/1.4 & 2.0 & 4.1/9.6 & $-1.0 \pm 0.6$ & $12.238\pm 0.020$ \\
\hline
\multirow{2}{*}{2015/02/02} & IRDIS H2H3 & 32x8x14 & \multirow{2}{*}{-9.3;39.4}  & \multirow{2}{*}{0.6;0.7} & \multirow{2}{*}{11} & \multirow{2}{*}{3} & $-134.155 \pm 0.006$  & $12.257 \pm 0.03$  \\
%\hline
 & IFS YJ & 64x4x9 & &  & & &  $-33.65\tablefootmark{e}  \pm 0.13$ & $7.46 \pm 0.02$   \\
\hline
\end{tabular}
\tablefoot{
\tablefoottext{a}{DIT is the individual detector integration time}
\tablefoottext{b}{Parallactic angle at the start and end of the observations. For the April 2014 observations, not all the available field rotation was used (see Section \ref{sec_obs} for details).}
\tablefoottext{c}{The coherence time $\tau_0$ is defined as $\tau_0=0.31r_0/v$, where $r_0$ is the Fried parameter measured by the DIMM and $v$ is the maximum of the wind speed measured at 30m height, and 0.4 times the predicted wind speed at an altitude of 400mbar \citep{Sarazin2002}.}
%\tablefoottext{c}{The DIMM coherence time is the greater value of either the Fried parameter $r_0$ divided by the wind speed at 30m or $r_0$ divided by the predicted wind speed (meteorogical models) at high altitude (where the pressure is 200 mbar) multiplied by 0.4 \citep{Sarazin2002}.}
\tablefoottext{d}{The calibration of the true north and platescale are detailed in \citet{Maire2015}. The true north indicated here includes the offset between the pupil-stabilised and field-stabilised mode.} 
\tablefoottext{e}{A rotation offset of $-100.46\pm0.13^\circ$ has been measured between IRDIS and the IFS.}
}
\end{table*}

\subsection{Data reduction}

\begin{figure*}
        \centering
        \includegraphics[width=\hsize]{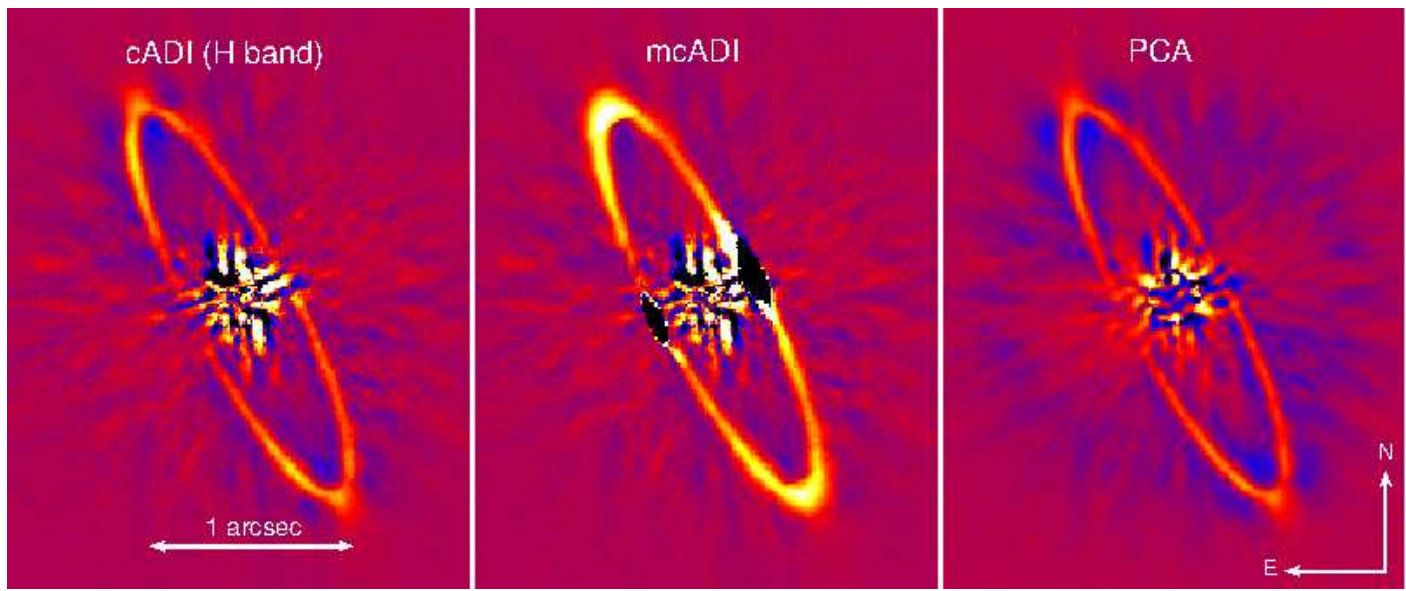}
        \caption{Images of the disc around HR\,4796\,A from IRDIS in the H-band, reduced with three different reduction algorithms: cADI, masked cADI and PCA (first five eigenmodes removed). North is up, east to the left. The colour scale is identical for all three reductions. The black region along the semi-minor axis of the mcADI image corresponds to region were the disc is entirely self-subracted therefore no information can be retrieved.}
        \label{fig_cADI_cADI-disc_PCA}
\end{figure*}

 We describe below the reduction performed on the IRDIS broadband H data, the IRDIS dual-band H2H3 data and the IFS data. For the IRDIS broadband H data obtained in 2014, the atmospheric conditions degraded in the course of the observations, this is why a severe frame selection was necessary to remove the bad frames. Under good adaptive optics correction, the disc of HR\,4796\,A is visible in a single DIT in the raw image. The data editing was performed by inspecting visually the raw frames and 74\% of frames were removed (out of the complete 42 min sequence). The raw frames were sky-subtracted, flat-fielded and bad-pixel corrected using the SPHERE data reduction and handling (DRH) pipeline \citep{Pavlov2008}. This set of frames is referred to as a cube, the third dimension being the time. The processed cubes were thereafter re-centred using the four satellite spots imprinted in the image during the centring sequence. With broadband filters, these satellite spots are elongated and we used the technique described in \citet{Pueyo2015} based on a Radon transform to determine the star location\footnote{We used the Radon-based centring technique developed in the Vortex Image Processing pipeline \cite[VIP,][available at \href{https://github.com/vortex-exoplanet/VIP}{https://github.com/vortex-exoplanet/VIP}]{Gomez2017}}. We checked that a visual adjustment of two lines passing through each opposite satellite spots agrees with the retrieved star location. We estimate the absolute centring accuracy to 0.5px or $\sim6$ mas. The individual images were not recentred because an active centring using the SPHERE differential tip/tilt sensor is dealing with the relative frame-to-frame centring \citep{Baudoz2010}. Three reduction algorithms were used: classical ADI \citep[cADI,][]{Marois2006}, masked classical ADI \cite[mcADI,][]{Milli2012} and Principal Component Analysis \citep[PCA,][]{Soummer2012,Amara2012}, shown in Fig. \ref{fig_cADI_cADI-disc_PCA}. The mcADI proceeds in two steps: a binary mask is first applied to the cube of pupil-stabilised images to mask in each frame the pixels corresponding to the ring. Because the disc rotates in the cube, the binary mask follows this rotation. We computed the median of this masked cube to build a reference coronagraphic image. In a second step, this reference coronagraphic image is subtracted from the unmasked cube and the cube is re-aligned and stacked. 
Because the atmospheric conditions were variable, the starlight leaking out of the coronagraph shows strong intensity variations. In order to better account for this variability in the star subtraction procedure of the cADI and mcADI algorithms, we introduced a scaling factor to weight the contribution of each frame in the reference coronagraphic image to be subtracted. This turned out to improve the level of residuals of the final reduced image by scaling down the contribution of the images where a lot of flux leaked out of the coronagraph. Each frame $i$ of the cube is divided by a factor $\lambda_i$, subtracted by the median of the resulting cube of renormalised images and then re-multiplied by $\lambda_i$ in order to preserve the photometry of the disc. The factor $\lambda_i$ is the total flux of frame $i$ within $1.75\arcsec$. The cube is then de-rotated and median-combined in order to obtain the final reduced image. For all three reductions, both \text{IRDIS} channels were combined to increase the signal-to-noise ratio (S/N).

\begin{figure*}
        \centering
        \includegraphics[width=\hsize]{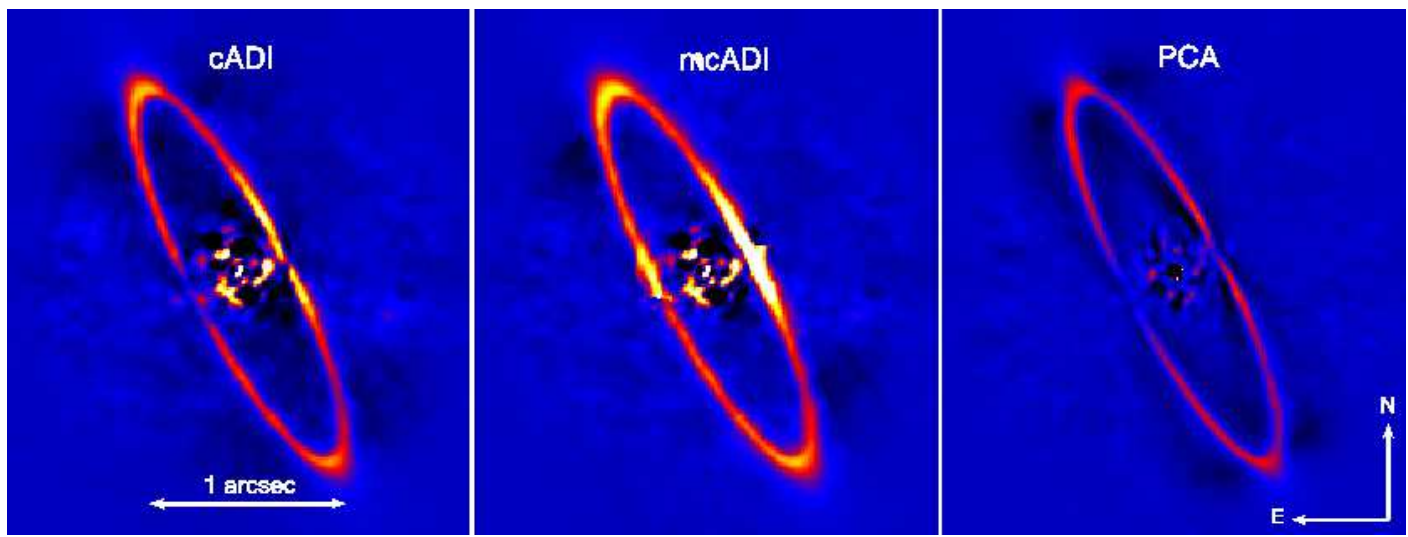}
        \caption{Images of the disc around HR\,4796\,A from IRDIS in the H2H3 filter, reduced with three different reduction algorithms: cADI, masked cADI and PCA (first five eigenmodes removed). The colour scale is identical. }
        \label{fig_irdisH23_cADI_mcADI_PCA}
\end{figure*}

\begin{figure}
        \centering
        \includegraphics[width=\hsize]{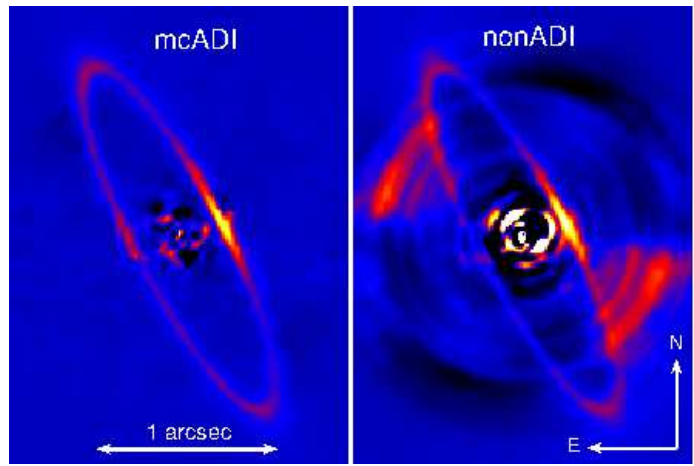}
        \caption{Masked cADI and non-ADI IRDIS H2H3 image, with a colour scale larger by a factor of ten with respect to Fig. \ref{fig_irdisH23_cADI_mcADI_PCA} to enhance the large dynamical range of the image between the very bright semi-minor axis in the west and the rest of the ring.}
        \label{fig_irdisH23_mcADI_nonADI}
\end{figure}

For the 2015 IRDIS dual-band H2H3 data, we also used the DRH pipeline for the standard cosmetic correction, and then performed four Gaussian fits on the satellite spots to determine the star centre behind the coronagraphic mask. We estimated the accuracy of the centring to 0.25px or 3mas. We applied similar reduction algorithms as for the 2014 data set (without requiring any renormalization here due to stable conditions), namely, cADI, mcADI and PCA, as shown in Fig. \ref{fig_irdisH23_cADI_mcADI_PCA}. The H2 and H3 filters were combined in order to increase the S/N, as no significant variations were noticeable between the two images. The more stable conditions, in particular the long coherence time, resulted in smaller starlight residuals close to the coronagraphic mask, revealing unambiguously for the first time the entire ring. %We favoured the mcADI reduction rather than the cADI-disc as for the 2014 data because it requires only priors on the shape of the disc and not on its intensity and is therefore more conservative.
To enhance the dynamic range of the mcADI image, we have displayed it on Fig. \ref{fig_irdisH23_mcADI_nonADI} with an unsaturated colour scale showing the full range of disc brightness. The stability of this data set allowed us to avoid resorting to ADI to detect the disc, enabling access to an unbiased view of the disc, free from ADI artifacts \citep{Milli2012}. This is shown in Fig. \ref{fig_irdisH23_mcADI_nonADI} (right). A simple azimuthal median was subtracted from each individual frame of the cube before de-rotating and stacking the cube. The two features extending $45^\circ$ counter-clockwise from near the ring ansae are instrumental artefacts: these are two brighter regions at the edge of the well-corrected area produced by a periodic pattern on the deformable mirror. This azimuthal asymmetry is totally subtracted in ADI but it is not removed by a non-ADI reduction. The de-rotation of the images smears this brighter region over an arc whose azimuthal extent equals the parallactic angle variation, as visible in the diagonal of Fig. \ref{fig_irdisH23_mcADI_nonADI} (right-hand panel). This does not, however, impede the analysis on the other part of the image, and confirms the view of the disc given by the mcADI image. 
%This procedure is not required for the PCA reduction because the data are automatically mean-subtracted before applying the Karuhnen-Lo\`eve transform.

\begin{figure*}
        \centering
        \includegraphics[width=\hsize]{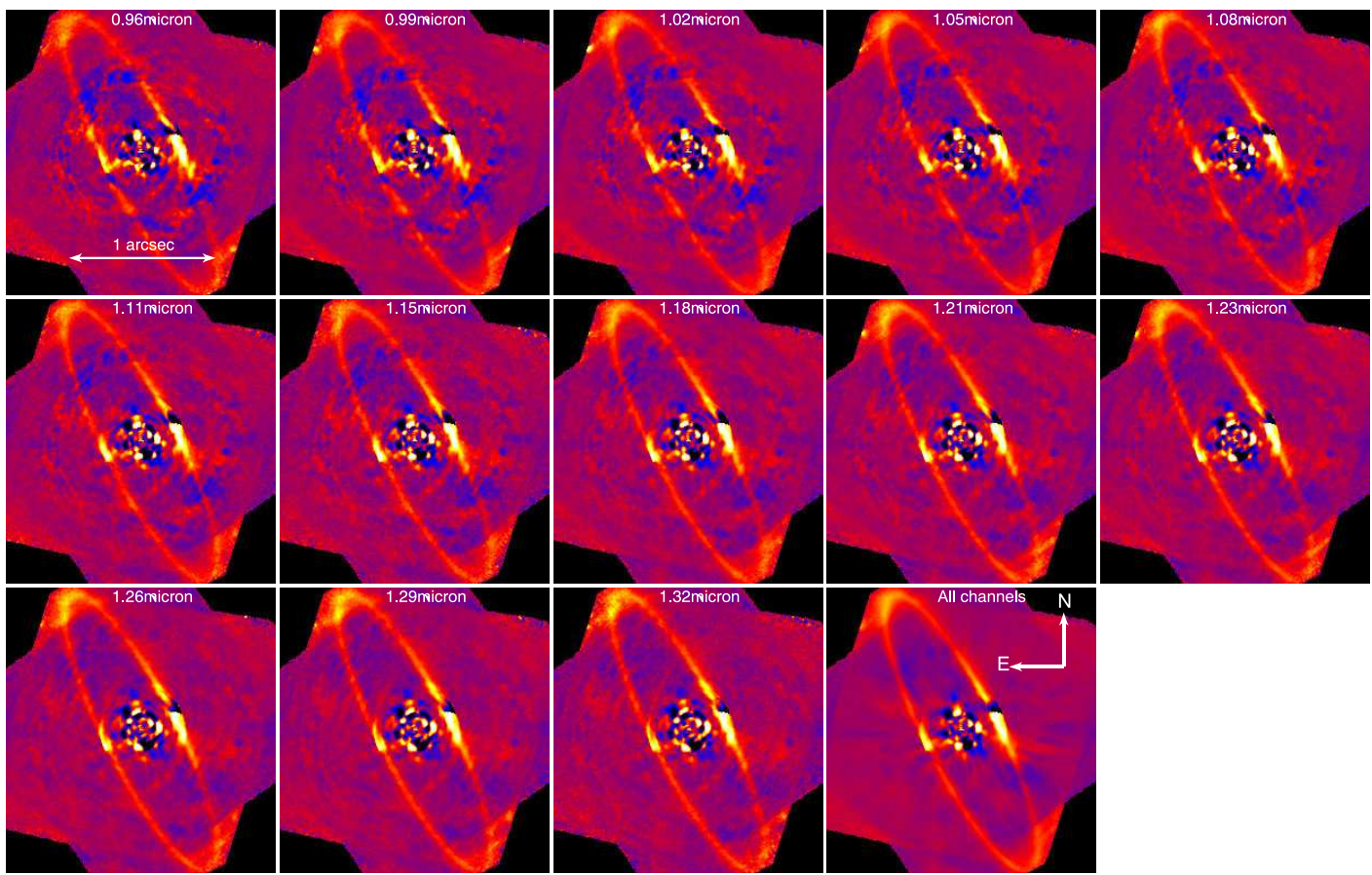}
        \caption{Masked cADI images obtained after binning three adjacent spectral channels of the IFS. The images were scaled by the flux of the PSF and the colour scale is identical for all images. The last image is the combination obtained by stacking all spectral channels of the IFS}
        \label{fig_ifs_mcADI_binned}
\end{figure*}

The IFS data were reduced using both custom routines and the DRH pipeline. The raw data were first sky-subtracted and bad-pixel corrected. A correction of cross-talk between spectral channels was applied to remove the high spatial frequency component of the cross-talk, as described in \citet{Vigan2015}. After building the master detector flat field, we called the DRH pipeline on arc lamp calibration data taken in the morning following the observations to associate each detector pixel with its corresponding wavelength and obtain a map called the pixel description table. The master flat field and pixel description table were used as input for the main science recipe of the DRH pipeline called sph\_ifs\_science\_dr, that builds the spectral cube consisting of 39 spectral channels and resamples each channel on a square regular grid of size 7.4 mas per pixel. The wavelength calibration was then more accurately determined using the arc lamp calibration files and the chromatic radial dependance of the  satellite spots, as described in detail in appendix A.2 of \citet{Vigan2015}. The spectral accuracy of this procedure is estimated to be 1.2 nm. For each spectral channel, the same three algorithms as those used to reduce the IRDIS images were applied, and the final images were normalised by the integrated flux within the central resolution element of the star  observed out of the coronagraphic mask. Fig. \ref{fig_ifs_mcADI_binned} (last panel) shows the IFS image averaged over all spectral channels and we provide in the other panels of Fig. \ref{fig_ifs_mcADI_binned} 13 normalised images obtained after mean-combining every three adjacent spectral channels. Because the disc diameter is slightly larger than the IFS field of view, the ansae are not visible during the whole sequence of observations, and the background noise is higher beyond 1.7\arcsec, for example in the ansae of the disc. Moreover, the disc being offset from the star towards the south-west (SW), the SW ansa spends a larger amount of time outside the IFS field of view than the north-east (NE) ansa, resulting in an apparent lower S/N.

\subsection{Contrast and planet detection limits}

\begin{figure}
        \centering
        \includegraphics[width=\hsize]{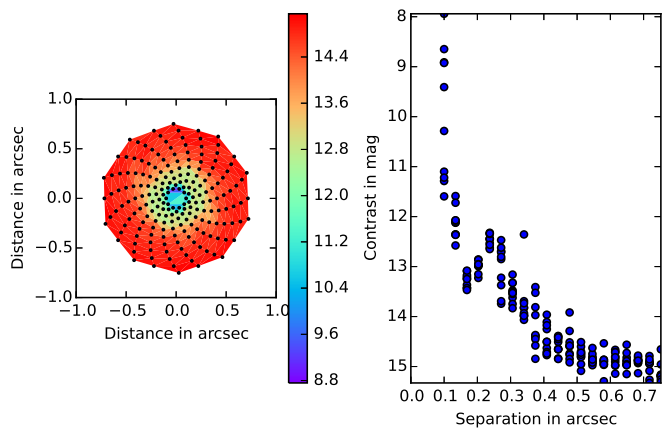}
        \caption{Left: Map of the $5\sigma$ detection limits expressed in magnitude for the IFS. The detection limits were computed at the position of the black dots, by injecting fake planets as detailed in \citet{Vigan2015}. Right: Radial curve of the detection limits. Each point corresponds to a fake planet in the left image. The contrast is independent of the wavelength because we assumed a stellar spectrum.}%The underlying map was created by using a Delaunay triangulation.
        \label{fig_detlim_mass_ifs}
\end{figure}

The derivation of the detection limits for the IFS data was done following the methodology described in \citet{Vigan2015}. We summarize here the main steps. To detect point sources, both angular and spectral differential imaging are used here. The images are first rescaled spatially so that the speckle pattern matches at all wavelengths. This leads to a rescaled cube of both spectral and temporal images, where the signal of a potential companion would move both with time and wavelength. This cube is processed using a PCA algorithm that subtracts from 1 to 500 modes in steps of ten. The same process is applied to the original cube of images where fake companions have initially been injected, in order to retrieve the S/N level of each companion in the reduced image. 
Fake planets are injected at separations from 100 mas to 750 mas on a spiral pattern to avoid any spatial or spectral overlap during the speckle subtraction algorithm. To properly sample the whole field, the analysis is repeated with the fake planets map injected at ten distinct orientations. The position of all injected fake planets is illustrated in Fig. \ref{fig_detlim_mass_ifs} left. The S/N is defined as the maximum pixel value of the image at the known location of the planet after convolution with a kernel of one resolution element size, divided by the rms of statistically independent pixels in an annulus located at the same separation as the planet. The penalty term from the small sample statistics described by \citet{Mawet2014} is taken into account. This process is repeated until a S/N of 5 is reached, the corresponding contrast in magnitude is shown in Figure \ref{fig_detlim_mass_ifs}. The fake planets are injected with the spectra of the central star HR\,4796\,A, for example with a constant contrast with respect to the star at each wavelength, which is a conservative assumption because spectral self-subtraction degrades the detection limits. An average contrast of 15 magnitudes is reached at 0.7\arcsec{} close to the edge of the field of view, and a value of 13 is obtained at 0.2\arcsec.

\begin{figure}
        \centering
        \includegraphics[width=\hsize]{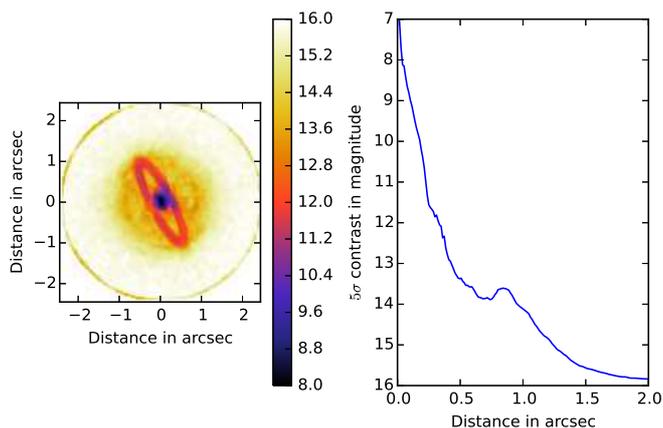}
        \caption{Left: Map of the $5\sigma$ detection limits expressed in magnitude for IRDIS in the H2 band. Right: Azimuthal median of the 2d contrast map displayed on the left. The bump at 0.8\arcsec{} is the limit of the correction radius of the adaptive optics system.}
        \label{fig_detlim_mass_irdis}
\end{figure}

For IRDIS, the detection limits are computed individually for each spectral channel by reducing the data using a PCA algorithm removing one single mode over the whole image from 0.02\arcsec{} to 2.4\arcsec. The flux losses due to the ADI process are also computed using fake planets injected at increasing radii in three branches at a level of about $5\sigma$. The contrast map shown in Fig. \ref{fig_detlim_mass_irdis} for H2 is defined as the rms in a box of $3\times3$ resolution elements, corrected for the flux losses and the small sample statistics. The contrast map for the H3 channel is almost identical. A constrast of 16 magnitudes is reached outside the correction radius of the adaptive optics system at 2\arcsec, and a value of 13.2 is obtained at 0.5\arcsec. 

The detection limits in contrast were converted in mass using the  AMES-Cond-2000 evolutionary model \citep{Allard2011}, assuming an age of 10 Myrs for the system \citep{Bell2015}. For the IFS, we used the mass to luminosity relation in broadband J, as validated by \citet{Vigan2015}. They indeed performed a detailed derivation of the detection limits for another A star, Sirius, based on the injection of fake planets using planetary atmospheric models. They showed that the results are well approximated by using a stellar spectra for the fake planets, for example, a constant contrast between the star and the planet throughout the IFS wavelength range, if one uses the longest IFS wavelength, the J band in our case, to convert the planet luminsoity in mass. The IFS observations are deeper than IRDIS and probe less massive planets below 0.4\arcsec, they are equivalent between 0.4\arcsec{} and 0.5\arcsec, and IRDIS is slightly deeper above 0.5\arcsec. A comparison with the best existing detection limits on the system, obtained with NaCo in the \Lp{} band is shown in Table \ref{tab_detlim_mass} for a few separations. We discuss the presence of planets based on these detection limits in Section \ref{sec_origin_offset_sharp} and also show there more specific radial curves of the detection limits along the semi-major and semi-minor axis of the disc. 

\begin{table}
\caption{Detection limits in Jupiter masses from these observation compared to the previous best limits set on the system by VLT/NaCo in the \Lp{} band \citep{Lagrange2012}, comparable to that of \citet{Rodigas2014} obtained at \Lp with MagAO/Clio-2.}
\label{tab_detlim_mass}
\begin{tabular}{c | c c }
\hline 
Separation in \arcsec & VLT/NaCo\tablefootmark{a} & VLT/SPHERE\tablefootmark{b} \\
\hline
0.1  & 32 & 15  \\
0.2  & 32 & 5  \\
0.5  & 3.5 & 2  \\
1.0  & 3.0 & 1.5  \\
1.5  & 2.0 & 1.5  \\
\hline
\end{tabular}
\tablefoot{
\tablefoottext{a}{The detection limits are from the Sparse Aperture Mask mode of NaCo \citep{Lacour2011} below 0.3\arcsec{} and from classical imaging above 0.3\arcsec.}
\tablefoottext{b}{The detection limits are from the IFS below 0.4\arcsec{} and from IRDIS above 0.4\arcsec.}
}
\end{table}

\section{Observed disc morphology}
\label{sec_morphology}

All reductions (Fig. \ref{fig_cADI_cADI-disc_PCA} to \ref{fig_ifs_mcADI_binned}) reveal the disc with a high signal to noise. The 2015 data show the entire ring, even the semi-minor axis which has, so far, always been hidden by strong starlight residuals at such a short separation ($\sim0.24$\arcsec). The 2014 data show a stronger S/N in the ansae, due to the wider spectral bandpass, but suffers from increased noise at short separations due to the poorer atmospheric conditions. Strong residuals from the diffraction by the four spiders of the telescope are indeed visible within 0.25\arcsec{} in Fig. \ref{fig_cADI_cADI-disc_PCA}.

\subsection{Surface brigthness radial profiles}

\begin{figure}
        \centering
        \includegraphics[width=\hsize]{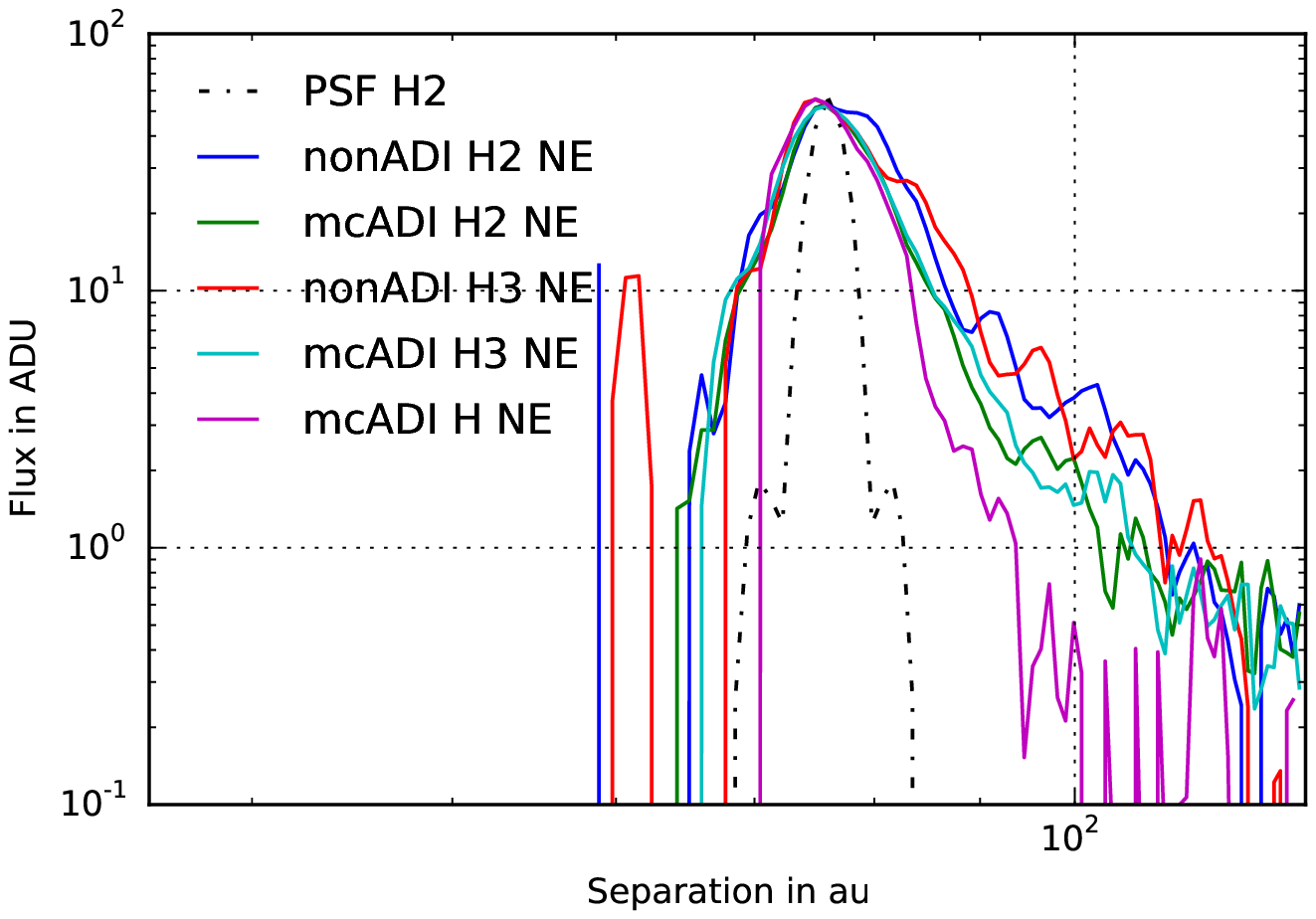}
        \includegraphics[width=\hsize]{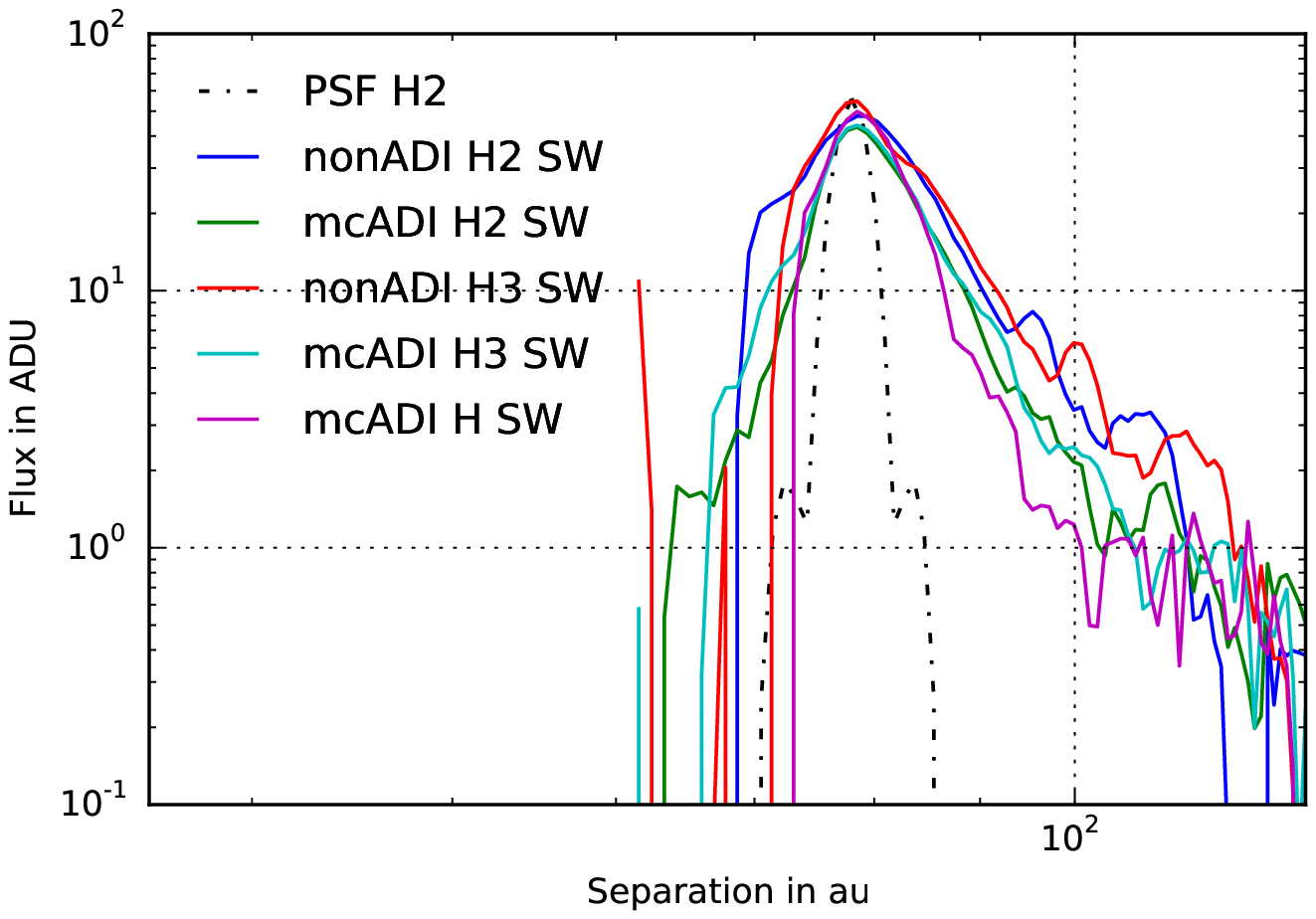}
        \caption{Radial profiles of the disc along the semi-major axis (NE ansa at the top, SW ansa at the bottom), shown here for different reductions, and filters. For comparison we overplotted the mean radial profile of the measured PSF in the H2 filter. The profiles have not been normalised but have by chance a similar flux in ADU for the filters shown here.}
        \label{fig_radial_profiles}
\end{figure}

The disc appears as a thin elliptical ring. It is clearly resolved radially both with IRDIS in the H or H2H3 filter and with the IFS in the Y band. 
We measured the radial profile of the disc at regular intervals along the ellipse. The profiles along the semi-major axis  for the mcADI and non-ADI reductions provide the least biased measurements of the ring true width. We show them in Fig. \ref{fig_radial_profiles}. The full width at half-maximum (FWHM) of the disc measured along the ansae is 0.12\arcsec{} at H2, compared to a FWHM of 0.046\arcsec{} for the PSF. The PSF profile has been overplotted in Fig. \ref{fig_radial_profiles} to illustrate this result.

Measurements of the true width of the ring are critical because they can be used to constrain the dust confinement mechanism \citep{Mustill2012}. \citet{Schneider2009} measured a value of $0.184\arcsec \pm0. 01\arcsec$ in a very wide-band in the optical after correcting for the width broadening effect of the HST~/~STIS PSF. In the near-infrared, ground-based measurements showed smaller values of $0.11\arcsec \pm 0.01\arcsec$  in the K band \citep{Perrin2015}, $0.102\arcsec$ in the H band \citep{Wahhaj2014}, $0.14\arcsec \pm 0.03\arcsec$ from 1 to 4\micron{} \citep{Rodigas2012} and $<0.14\arcsec{}$ in the L band \citep{Lagrange2012}, all after correction for the PSF convolution.

Two effects mainly affect the measured width of the ring: the convolution with the PSF and the potential bias from the reduction technique. We introduced a fake disc in the raw images at $90^\circ$ from the real one, reduced the images with mcADI and non-ADI again, and performed the width measurements on both ansae of the fake disc. We found, as also shown in \citet{Lagrange2012}, that the most important effect is due to the PSF convolution, which increases the width by 26\% and that mcADI does not bias the measured width with respect to the width of the convolved disc to more than 1\%. In nonADI, the images suffer from a high residual noise, hence a larger dispersion. 

The FWHM after correction for the PSF convolution is displayed in Table \ref{tab_radial_prof}. To compute the error bar, we assigned an error to each pixel of the ansa radial profile, defined as the standard deviation of the pixels within an annulus at the same radius (after masking the ring). We then measured again the FWHM after adding gaussian noise to the ansa profile and repeated this process $10^4$ times to estimate the dispersion. Because of the low-frequency noise in the non-ADI images, this uncertainty appears large but it is fully consistent with the mcADI values, and one has to bear in mind that this is the first time such a measurement is possible on an image without performing any star-subtraction algorithm. The ring width is  narrower than that of the STIS data and we suspect that this arises from both a systematic bias and a physical effect. Indeed, a pure physical effect with a ring wider in the optical might be expected if small grains,  which are less efficient near-infrared scatterers, are being blown out of the system. In this case, the outer half-width at half-maximum (HWHM) is expected to be wider in the optical. We however measured that both the inner and outer HWHM are smaller with IRDIS in the near-infrared than with STIS in the optical. Therefore, a physical effect is not enough to explain this discrepancy. It is however likely to play a minor contribution, because the discrepancy is smaller for the outer HWHM. We think that the main explanation comes from a systematic bias between the two measurements. In particular we suspect that the simple quadratic subtraction used to correct for the PSF convolution with STIS underestimates the intrinsic FWHM of the ring due to the very steep inner and outer profiles, as detailed below.

\begin{table}
\caption{Ring radial FWHM along the semi-major axis. The uncertainty is given at $3\sigma$.}
\label{tab_radial_prof}
%\centering
\begin{tabular}{c  c c | c c }
%\begin{tabular}{p{1.5cm} p{2cm} p{2cm} }
\hline 
Data set & Side & Unit & mcADI & non-ADI \\
\hline
\hline
\multirow{2}{*}{IRDIS H2}& \multirow{2}{*}{NE} & \arcsec{}  & $0.092 \pm 0.011$ & $0.111 \pm 0.043$  \\
                                       &                                & au            & $6.7 \pm 0.8$ & $8.1 \pm 3.1$  \\
\hline
\multirow{2}{*}{IRDIS H2}& \multirow{2}{*}{SW} & \arcsec{}  & $0.096 \pm 0.014$ & $0.137 \pm 0.050$  \\
                                       &                                & au            &  $7.0 \pm 1.0$ & $9.9 \pm 3.6$  \\

\hline
\multirow{2}{*}{IRDIS H3}& \multirow{2}{*}{NE} & \arcsec{}  & $0.099 \pm 0.014$ & $0.099 \pm 0.050$  \\
                                       &                                & au            & $7.2 \pm 1.0$ & $7.2 \pm 3.6$  \\

\hline
\multirow{2}{*}{IRDIS H3}& \multirow{2}{*}{SW} & \arcsec{}  & $0.099 \pm 0.013$ & $0.123 \pm 0.113$ \\
                                       &                                & au            & $7.2 \pm 1.0$ & $9.0 \pm 8.2$  \\

\hline
\multirow{2}{*}{IRDIS H}& \multirow{2}{*}{NE} & \arcsec{}   &  $0.096 \pm 0.011$ & NA  \\
                                       &                                & au            & $7.0 \pm 0.8$ & NA  \\

\hline
\multirow{2}{*}{IRDIS H}& \multirow{2}{*}{SW} & \arcsec{}    & $0.088 \pm 0.014$ & NA \\
                                       &                                & au            & $6.4 \pm 1.0$ & NA \\
\hline
\end{tabular}
\end{table}

\begin{figure}
        \centering
        \includegraphics[width=\hsize]{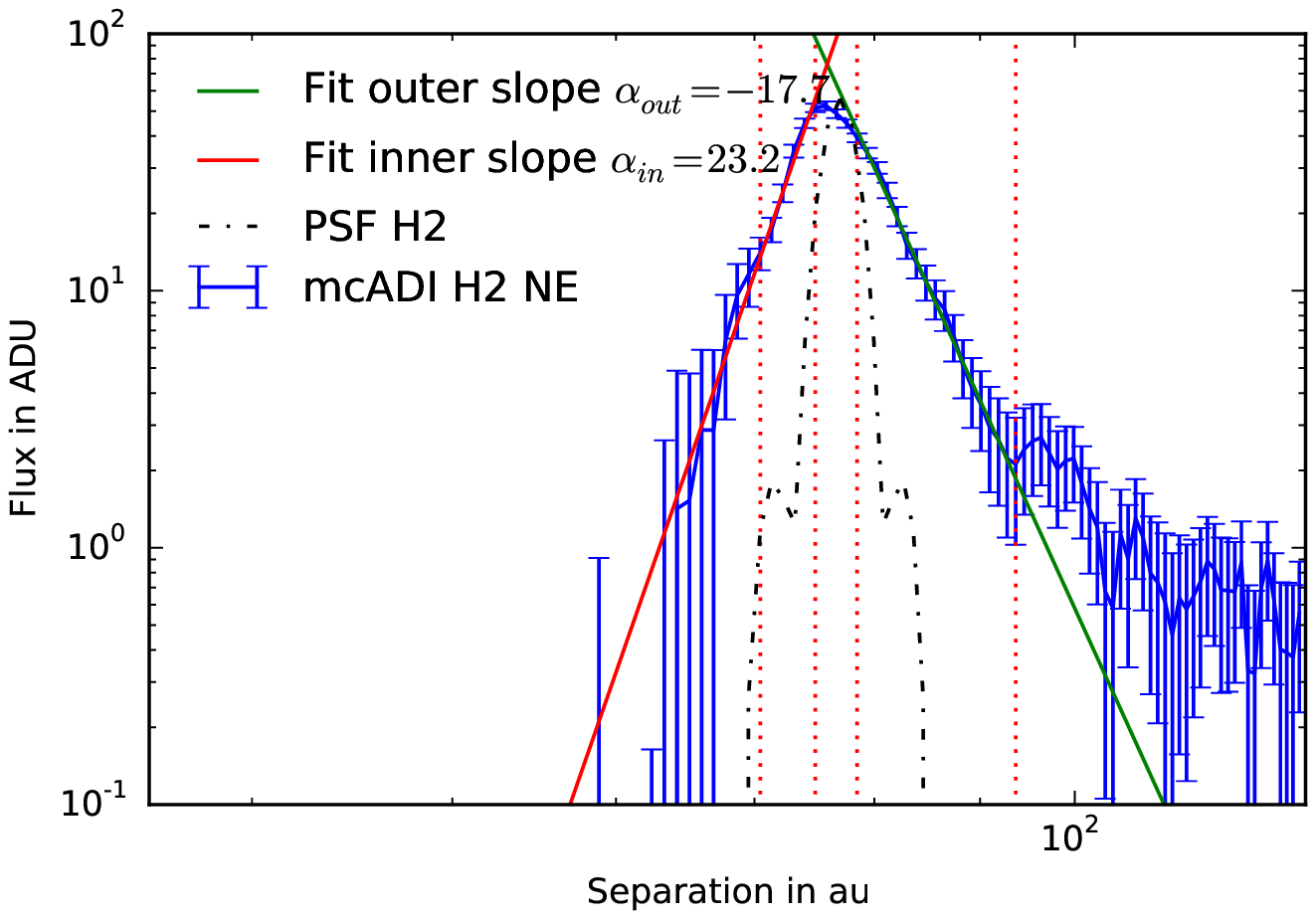}
        \includegraphics[width=\hsize]{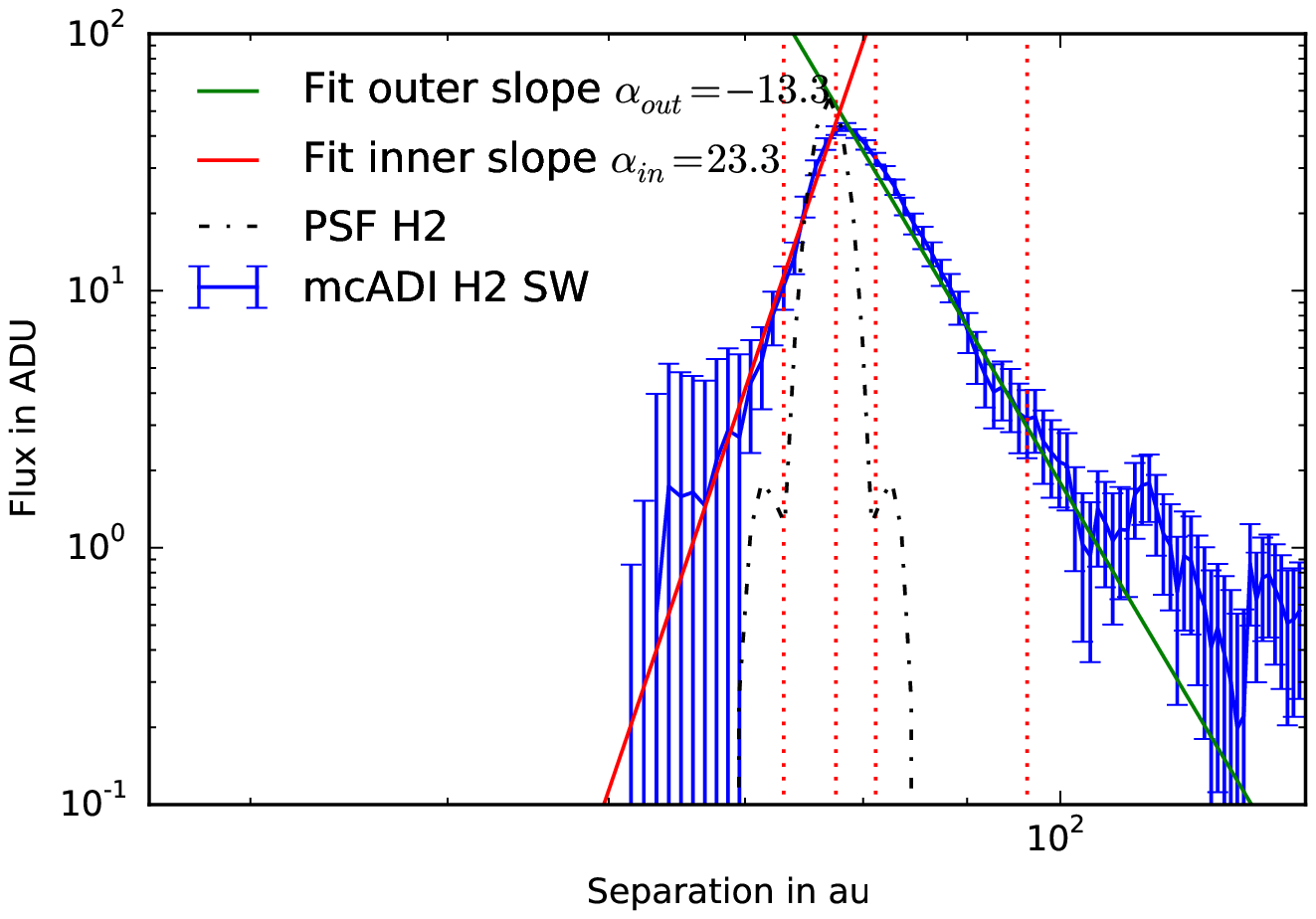}
        \caption{Radial profiles along the semi-major axis of the disc as measured in the H2, mcADI-reduced image. The vertical red dotted lines show the boundary used for the fit of the inner and outer profile with a power law. The profile of the PSF is indicated as a reference. The top image corresponds to the NE ansa and the bottom one to the SW ansa.}
        \label{fig_mcadi_slope_measurement}
\end{figure}

The disc profile is asymmetric, with a slope steeper inside than outside. To quantify this asymmetry, we fitted a power law of equation $\Lambda \times r^{-\alpha}$ to the inner and outer radial profile. We measured the inner slope $\alpha_{in}$ over 0.06\arcsec{} or 4.5 au, prior to the peak brighntess of the ring, and the outer slope $\alpha_{out}$ over 0.21\arcsec{} or 15 au, after the peak. Figure \ref{fig_mcadi_slope_measurement} illustrates this measurement for the H2 image reduced with mcADI. The measurement is sensitive to the boundaries used for the fit. For homogeneity of the measurements presented here, we have used the same boundaries for all the images where the fit was performed (different filters and reduction techniques). For the inner profile, we cannot use regions at more than 73 mas from the peak towards the star, either because of self-subtraction in case of the mcADI reduction or because of strong starlight residuals in non-ADI. For the outer profile, we limited the fitting area to regions within 0.25\arcsec{} of the peak, as shown in Fig. \ref{fig_mcadi_slope_measurement}.

\begin{figure}
        \centering
        \includegraphics[width=\hsize]{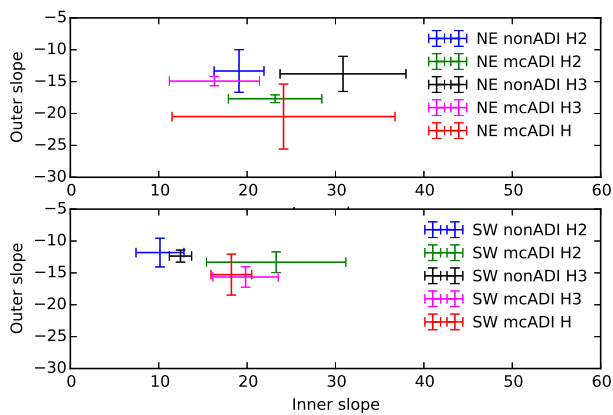}
        \caption{Inner and outer slope of radial brightness profile along the semi-major axis (NE ansa on the top, SW ansa on the bottom), with a $3\sigma$ error bar.}
        \label{fig_ain_aout}
\end{figure}

The measured slopes are displayed in Fig. \ref{fig_ain_aout}. The different measurements display a large uncertainty but are overall compatible and show that the disc displays an overall inner slope $\alpha_{in}=18\pm3.5$ and an outer slope of $\alpha_{out}=-13\pm2.3$ (mean of the non ADI measurements, least biased by the reduction technique). The inner slope is very steep but not as steep as the slope of the measured PSF (see Fig. \ref{fig_radial_profiles}). As an exercise, we  modelled a disc with a sharp step-like transition for the inner and outer edges, and measured after convolution and mcADI reduction an inner slope of $35\pm1.4$ and an outer slope of $-32\pm1.8$. The measurements of Fig. \ref{fig_ain_aout} are therefore not compatible with a sharp transition for the outer edge of the disc and only marginally compatible with a sharp inner edge.  

\subsection{Centre offset of the ring}

The ring is known to be offset from the star. Several authors previsouly measured the geometry of the ring using the maximum merit procedure described in \citet{Buenzli2010} and \citet{Thalmann2011}. 
Because the disc is now detected along all azimuths, we developed an alternative method based on a discrete sampling of the ring, which turns out to be more sensitive to the ellipse parameters. For a given azimuth, we fitted a smooth combination of two power laws described by the following equation initally introduced by \citet{Augereau1999} to the
radial profile of the image:

\begin{equation}
\label{eq_2powerlaw}
I(r) = I_0\times \left( \frac{2}{\left( \frac{r}{r_0} \right)^{-2\kappa_{in}} + \left( \frac{r}{r_0} \right)^{-2\kappa_{out}} }\right)^{1/2}
.\end{equation}
We derived the radius of the maximum brightness of the ring for that azimuth. We repeated this measurement for different azimuths in order to sample regularly the disc every resolution element (one FWHM). We then found the best ellipse passing through these points. An illustration of those measurement points along with the best ellipse is given in the top right hand image of Fig. \ref{fig_triangle_irdis_H}, \ref{fig_triangle_irdis_H2}, \ref{fig_triangle_irdis_H3} and \ref{fig_triangle_IFS} in Appendix \ref{App_MCMC}.
To find the best ellipse passing through the measurement points, we implemented the non-linear geometric fitting approach described in \citet{Ray2008}. We used a Markov chain Monte Carlo technique (hereafter MCMC) to find the best ellipse minimising Eq. 15 of \citet{Ray2008}. We chose to implement the MCMC with the affine-invariant ensemble sampler called emcee \citep{Foreman-Mackey2013}. By doing this, we retrieved the parameters of the projected ellipse in the plane of the sky: the projected semi-major axis $a^\prime$, the projected semi-minor axis $b^\prime$, the offsets $\Delta\alpha$ and $\Delta\delta$ in right ascension and declination of the ellipse centre with respect to the star location, and the position angle $PA$. These parameters are given in Table \ref{tab_ring_morphological}, in the rows corresponding to "projected ellipse", together with the uncertainty measured directly on the posterior probability density function of the fitted parameters. 
Using the Kowalsky deprojection technique described in \citet{Smart1930} for binary systems and also applied by \citet{Stark2014,Rodigas2015} on debris discs, we derived the parameters of the true ellipse described by the dust particles in the orbital plane: the true semi-major axis $a$, the eccentricity $e$, the inclination $i$, the argument of pericentre $\omega$ and the longitude of ascending node $\Omega$. This technique uses the exact same input as the direct elliptical fit and the same metric to compute the distance between a model and the measurements; the only difference being that the parameter space is the ellipse true orbital elements which are then converted in the sky plane before computing the likelihood of each model. The result is given in Table \ref{tab_ring_morphological}, in the rows corresponding to "deprojected ellipse".

As a sanity check of this new method introduced to measure the morphology of a ring, we fit a model disc directly from the reduced image, using the GraTeR code \citep{Augereau1999}, a procedure more similar to the maximum merit technique. This  gives a very good agreement with the new technique described above. The description and results of this sanity check are given in Appendix \ref{App_sanity_check}.

\begin{table*}
\caption{Projected and deprojected ring parameters. The error is given at a $3\sigma$ level and contains only the statistical error from the fit and no systematic error from the true north or star registration.}
\label{tab_ring_morphological}
\centering
\begin{tabular}{c c |c c c c c}
\hline 
\multicolumn{2}{c}{Type of fit} & Parameter &  IRDIS H & IRDIS H2 & IRDIS H3 & IFS  \\
\hline
\hline
\parbox[t]{2mm}{\multirow{5}{*}{\rotatebox[origin=c]{90}{Projected}}} & \parbox[t]{2mm}{\multirow{5}{*}{\rotatebox[origin=c]{90}{ellipse}}} &   $a^\prime$(mas) & $ 1064 \pm     6$ & $ 1064 \pm     8$ & $ 1066 \pm     8$ & $ 1059 \pm     4$  \\% $1065 \pm 6.9 $ \\
                                                 & &   $b^\prime$(mas) & $  252 \pm     4$   &  $  249 \pm     3$ & $  248 \pm     3$ &  $  249 \pm     2$ \\%& $249.5 \pm 3.0$ \\
                                                 & &   $\Delta\alpha$(mas) & $   -4 \pm     4$  &  $   -4 \pm     4$ & $   -3 \pm     4$ & $   -7 \pm     2$  \\%& $-4.5 \pm 3.6$ \\
                                                 & &   $\Delta\delta$(mas) & $  -28 \pm     5$  &  $  -23 \pm     6$ &  $  -23 \pm     6$ & $  -14 \pm     4$  \\%& $-22 \pm 5.3 $\\
                                                 & &   PA($^\circ$) &  $27.69 \pm  0.26$ & $27.00 \pm  0.25$ & $26.99 \pm  0.27$  & $26.81 \pm  0.16$  \\%& $27.14 \pm 0.24$\\
\hline
\parbox[t]{2mm}{\multirow{5}{*}{\rotatebox[origin=c]{90}{Deprojected}}} & \parbox[t]{2mm}{\multirow{5}{*}{\rotatebox[origin=c]{90}{ellipse}}}  & $a$(mas) & $ 1066 \pm     6$  & $ 1064 \pm     8$ & $ 1067 \pm     8$ & $ 1061 \pm     5$ \\
                                                   &   & $e$ & $0.070 \pm 0.011$ & $0.059 \pm 0.010$ &  $0.057 \pm 0.011$ & $0.052 \pm 0.007$   \\
                                                    &  & $i$($^\circ$)  & $76.33 \pm  0.24$ & $76.48 \pm  0.24$ & $76.55 \pm  0.24$ & $76.42 \pm  0.15$  \\
                                                     & & $\omega$($^\circ$) &  $-72.44\pm  5.10$ & $-73.03\pm  6.91$ &  $-71.60\pm  7.72$ & $-80.15\pm  4.44$ \\
                                                      & & $\Omega$($^\circ$) &  $27.71 \pm  0.25$ & $27.02 \pm  0.25$ &  $27.02 \pm  0.27$ & $26.82 \pm  0.16$ \\
\hline
\end{tabular}
\end{table*}

Moreover, ADI is known to introduce biases in the morphological parameters extracted from the reduced image \citep{Milli2012}. Here, the use of masked cADI does indeed minimise disc self subtraction and biases but it does not totally remove them. Therefore, we analysed these biases by repeating the measurement procedure described previously on a model disc image generated with known parameters. We used the best ellipse parameters ($a$,$e$,$i$,$\omega$,$\Omega$) as derived in Table \ref{tab_ring_morphological}, from the deprojeted image technique for the H2 band. We then inserted the disc model image in a fake pupil-stabilised cube with the same position angle as in the real H2 observations. We convolved each image by the PSF measured at H2, reduced the cube with the mcADI algorithm, and repeated the measurement procedure developed above to retrieve the disc parameters. We found that the bias from the PSF convolution and ADI data reduction on the semi-major axis $a$ of the disc is negligible (0.1\%), as well as on the inclination of the disc (deviation of less than $0.2^\circ$, smaller than the uncertainty). However the deviation is of the order of the uncertainty for  the eccentricity ($0.009$) and for the argument of pericentre $\omega$ ($8^\circ$), and there is a significant bias of $0.9^\circ$ on the PA of the line of nodes $\Omega$. We therefore include this systematic source of error in the final error bar given in Table \ref{tab_final_disc_param}.

The average of these measurements is summarised in the last column of Table \ref{tab_final_disc_param}, including all sources of errors.
These measurements show that the disc is elliptic with a mean ellipticity of $0.059 \pm 0.020$ (average for the "deprojected ellipse" technique), in good agreement with \citet{Rodigas2014} who estimated $0.060 \pm 0.020$. The argument of pericentre $\omega$ is $-74^\circ \pm 12^\circ$, which means that it is close to the semi-minor axis of the projected image of the disc, in the north-east quadrant. We note that the value of $\omega$ reported here is compatible with the Fig. 3 of \citet{Rodigas2014} but not with their numerical value of $110.6^\circ \pm 12.6^\circ$ and we suspect that the definition of $\omega$ for both articles differs by a factor $180^\circ$ because of the opposite assumption for the forward-scattering side. The inclination is compatible with previous measurements by \citet{Thalmann2011,Schneider2009} and \citet{Rodigas2014}

\begin{table}
\caption{Weighted-averaged disc deprojected parameters combining all bands. The uncertainty is given at $3\sigma$ and includes measurement uncertainties, systematics from the instrument and from the data reduction algorithm.}
\label{tab_final_disc_param}
\centering
\begin{tabular}{c c}
\hline 
$a$ (mas) & $1065 \pm 7$ \\
$e$ & $0.06 \pm 0.014 $ \\
$i$ ($^\circ$) & $76.45 \pm 0.7$ \\
$\omega$($^\circ$) & $-74.3 \pm   6.2$ \\
$\Omega$ ($^\circ$) & $27.1 \pm 0.7$ \\
\hline
\end{tabular}
\end{table}

%\subsection{Deprojected view of the disc}

\begin{figure}
        \centering
        \includegraphics[width=\hsize]{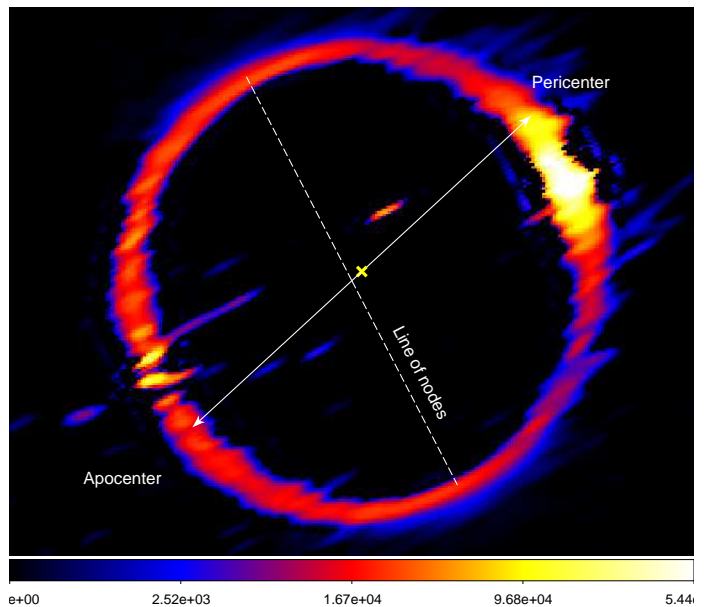}
        \caption{Deprojected view of the ring, after deconvolution of the H2 image. The colour scale is logarithmic, north is up, east to the left. The yellow cross indicates the location of the star.}
        \label{fig_deprojection}
\end{figure}

Figure \ref{fig_deprojection} shows the  deprojected image of the disc at H2, assuming the ring has no vertical thickness. 
Because the on-sky projected image is the original disc convolved by the PSF of the instrument, the image appears after deprojection convolved by an elliptical PSF, which biases our view of the disc. We therefore deconvolved the image prior to deproject it. We used the deconvolution algorithm MISTRAL \citep{Conan1999} adapted for adaptive optics images with imprecise knowledge of the PSF. The deprojected view enhances the bright asymmetry due to the anisotropic phase function of the disc, as already seen on the projected image. The brightest part of the ring appears also thicker. A possible explanation is the small but non-zero vertical height of the disc combined with a very anisotropic scattering phase function. It is also seen on model discs combining those two properties. Indeed, along the semi-minor axis towards the star, the scattering angle can be smaller than $13.6^\circ$ above the midplane, if the disc is not vertically flat. A very steep phase function could therefore compensate the smaller dust density away from the midplane to make the ring appear thicker towards the star. On the other hand, two regions appear fainter, in the north-west (NW) and south-west (SW), apart from the pericentre. This is probably physical and can originate from a dip in the scattering phase function of the dust, as discussed in the next section, or a decrease in the dust density close to the true semi-minor axis of the disc. We also note that at this SW position, previous observations tentatively showed a distortion in the ring \citep{Lagrange2012,Thalmann2011}, but these new observations do not confirm this feature. 
The deprojected image also shows blobs in the regions initially the closer to the star before the deprojection. They are probably artifacts resulting from the deconvolution, later elongated perpendicular to the line of nodes by the deprojection. The gaps seen in the SE ansa are probably not physical, and are related to the large flux losses from ADI occuring along the semi-minor axis of the disc (detailed later in Section \ref{sec_dust_prop}).  

\section{Observed dust scattering properties}
\label{sec_dust_prop}

With the new IRDIS observations, we can now probe the scattering phase function (hereafter SPF) at angles never accessible up to now. By increasing this range of scattering angles, we intend to confirm that what was interpreted in the past as a slight preferential forward scattering \cite[e.g.][]{Schneider2009} turns out to be a slight preferential backward scattering, with a peak of forward-scattering on the other side of the disc, as already proposed to explain recent scattered light observations \citep{Milli2015,Perrin2015}. These new conclusions enable us to reconcile the polarised and non-polarised images without the need for an optical depth about unity, as proposed by  \citet{Perrin2015}.

\subsection{Phase function of the disc}
\label{sec_phase_function}

\begin{figure}
        \centering
        \includegraphics[width=0.5\hsize]{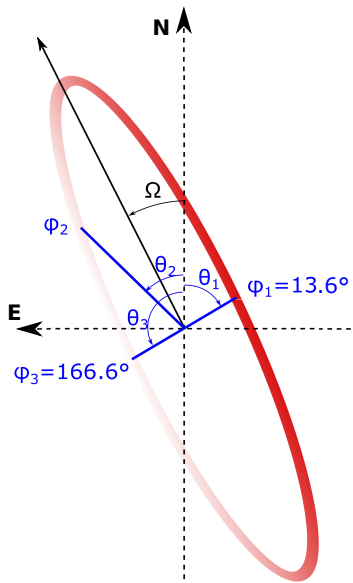}
        \caption{Schematics of the ring, defining the angles $\theta$ and $\phi$ used in Eq. \ref{eq_phi}. We plotted as an illustration 3 points along the ring, defined by the PA $\theta_i$ and corresponding to a scattering angle $\phi_i$}
        \label{fig_schematics_angle}
\end{figure}

Knowing the true orbital elements of the ring (Table \ref{tab_ring_morphological}), we can derive the SPF of the dust, as it was done for the debris disc around HD181327 \citep{Stark2014}.
The underlying assumptions are twofold. First we must assume that the disc has a negligible scale height with respect to the radial extension, so that each point along the ring corresponds to a unique value of the scattering angle. Second, we must also assume that the dust density distribution is uniform azimuthally and the dust properties are identical azimuthally. In other words, after correcting for the distance between the scatterers and the star, the ADI flux loss and the convolution by the PSF, any azimuthal brightness variation along the ring is entirely attributable to the shape of the SPF. The data are consistent with these two assumptions, as we shall see. To retrieve the SPF, we proceeded as follows: 

First, we regularly sampled the best ellipse (as defined in the first row of Table \ref{tab_ring_morphological}). The spacing between each point was set to one resolution element. We associated to each point at position angle $\theta$ in the plane of the sky a unique scattering phase angle $\varphi$ given by the following expression
\begin{equation}
\varphi = \text{arcsin}\left( \frac{1}{\sqrt{\sin^2(\theta-\Omega)/\cos^2i+\cos^2(\theta-\Omega)}} \right) 
\label{eq_phi}
.\end{equation}
We used in this expression the average inclination $i$ and average position angle of the line of nodes $\Omega$ from Table \ref{tab_final_disc_param}. We considered values of $\varphi$ between $0$ and $180^\circ$, assuming that the forward-scattering side of the disc (0$\leq\varphi\leq90^\circ$ ) is on the NW (see discussion below). A schematics illustrating those angles is shown in Fig. \ref{fig_schematics_angle}.

Second, for each location, we performed aperture photometry on the as-observed (projected) view of the ring, requiring therefore elliptical apertures to account for the projection effect. Each elliptical aperture were oriented along the PA of the disc, had a semi-major axis 0.1\arcsec{} (about the FWHM of the ring) and with the same major to minor axis ratio as the disc (i.e. $4.25)$. Using elliptical apertures with such an aspect ratio on the projected image is identical as using circular apertures on a deprojected image of the disc. With the former technique, the noise estimation is easier because it is only radially dependent on the on-sky projected image, whereas it also depends on the azimuth for a deprojected image.

\begin{figure}
        \centering
        \includegraphics[width=\hsize]{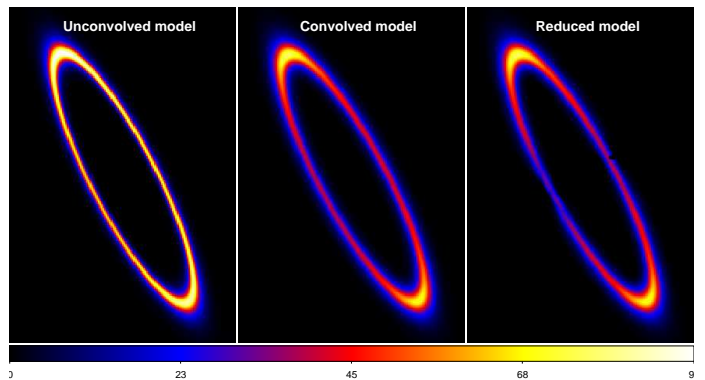}
        \caption{Disc images with the same colour scale showing the effect of the convolution and the mcADI reduction.}
        \label{fig_calib_reduction}
\end{figure}

\begin{figure}
        \centering
        \includegraphics[width=\hsize]{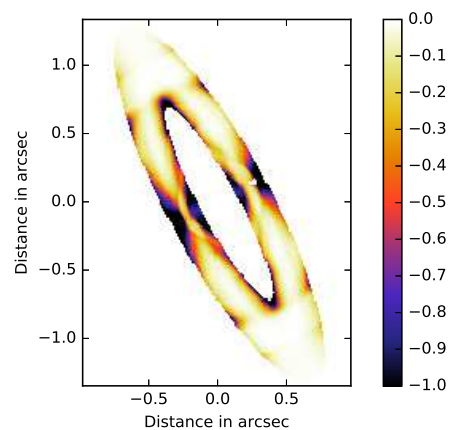}
        \caption{Map of the flux loss resulting from the mcADI reduction. A value of 0 indicates the absence of flux loss while a value of -1 means all the disc flux is removed by ADI.}
        \label{fig_flux_loss}
\end{figure}

Last, we corrected the flux measured in each aperture by three terms: the inverse physical distance squared due the stellar illumination, a correction term to account for ADI flux losses, and a correction term to accound for the convolution by the instrumental PSF. Those three terms depend on the position along the ring and therefore have an impact on the derived phase function. To compute the last two terms, we used an isotropic scattered light model of the disc created with the GrAteR code \citep{Augereau1999}, with the parameters described in Table \ref{tab_final_disc_param}, illustrated in Fig. \ref{fig_calib_reduction} left. We compared the elliptical aperture photometry of the inital unconvolved model disc with that of the final image after insertion of the model in a fake pupil-stabilised cube, convolution by the PSF and mcADI reduction. To do so, we inserted the model in a fake pupil-stabilised cube of images, with the same orientation as seen during the observations and each image was convolved by the PSF. Fig. \ref{fig_calib_reduction} middle shows this convolved model. The effect of the convolution is mainly to enhance the ansae. Then the cube is reduced using the mcADI algorithm (Fig. \ref{fig_calib_reduction} right). With the masking strategy, the ADI flux losses are minimised to less than 10\% in most areas of the disc, and affect mostly the semi-minor axis because the mask was slightly undersized with respect to the disc true width to avoid being unable to evaluate the reference coronagraphic image on a large region. A map of the flux loss is shown in Fig. \ref{fig_flux_loss}. 

\begin{figure}
        \centering
        \includegraphics[width=\hsize]{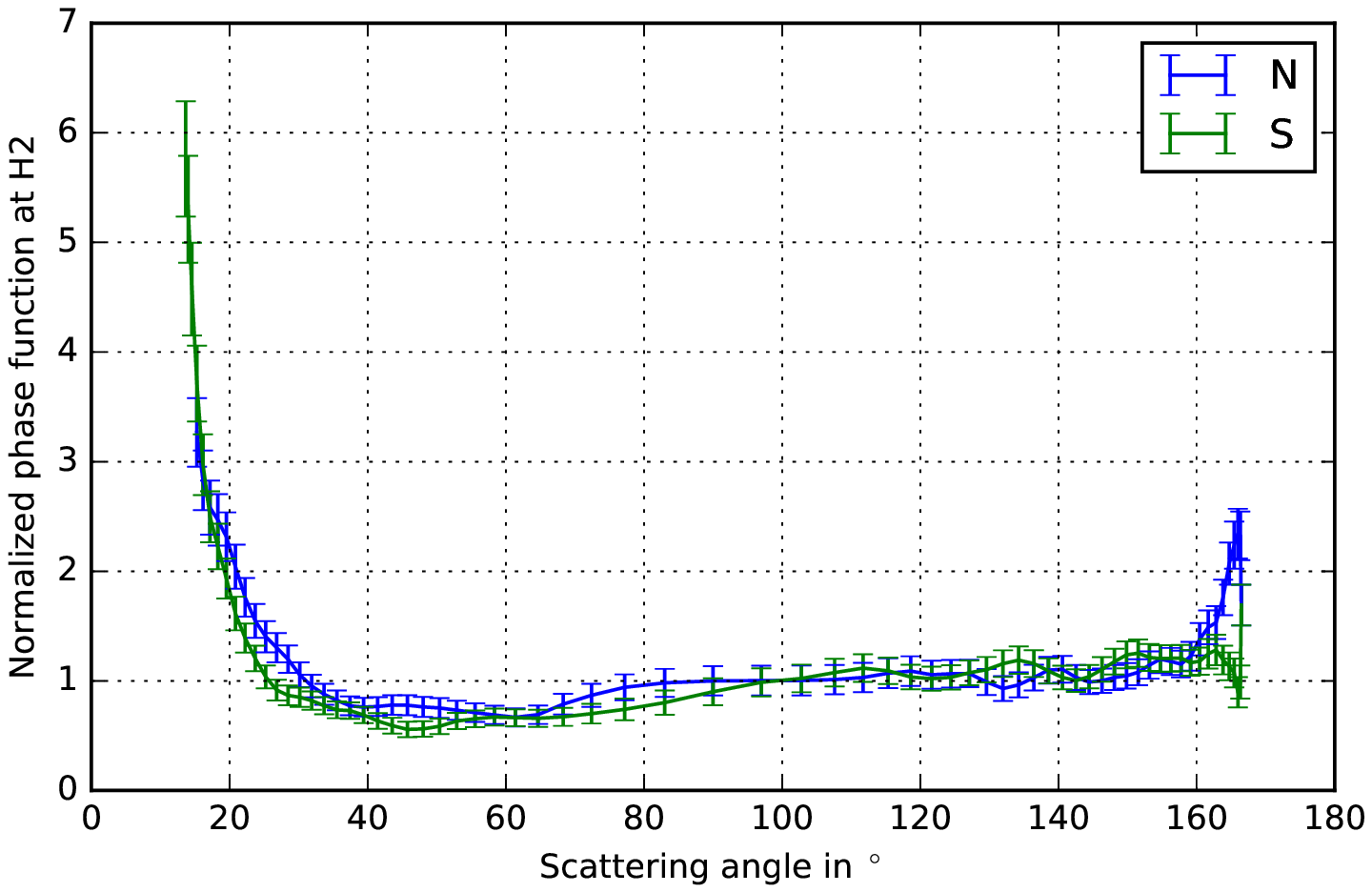}
        \includegraphics[width=\hsize]{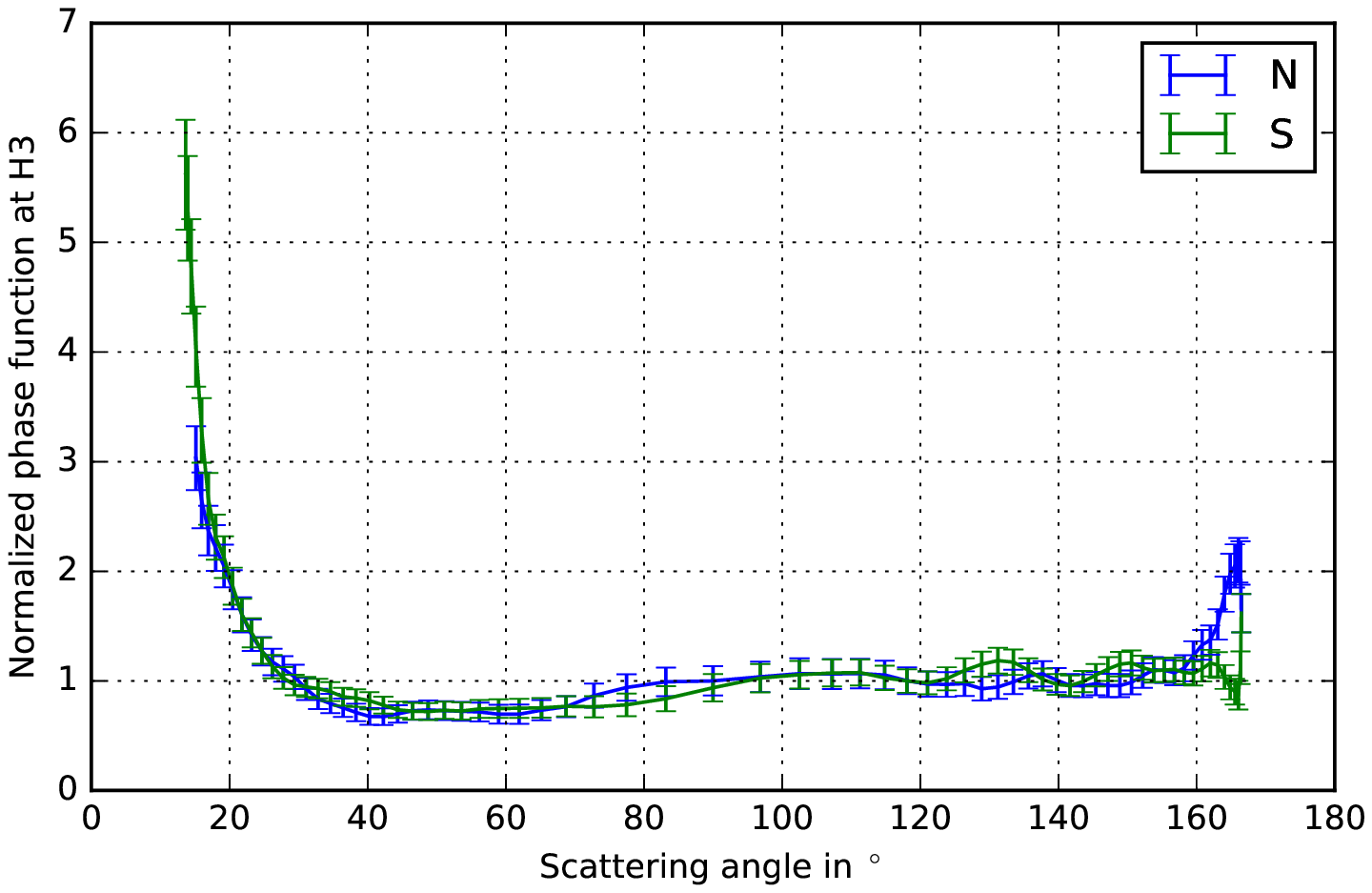}
        \caption{Phase function of the dust (H2 at the top, H3 at the bottom) normalised for the north side at $90^\circ$. The uncertainty is given at a $3\sigma$ level.}
        \label{fig_phase_function}
\end{figure}

\begin{figure}
        \centering
        \includegraphics[width=\hsize]{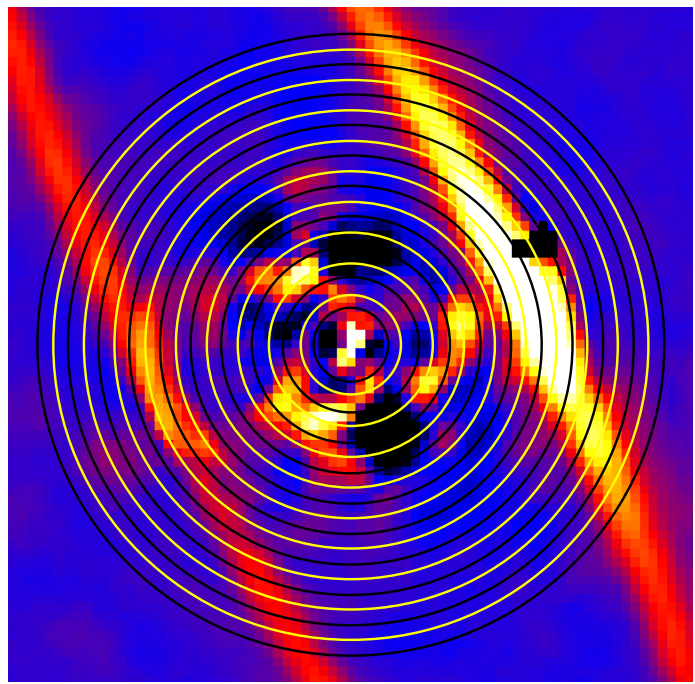}
        \caption{Position of the bright (yellow circles) and dark (black circles) Airy rings from the diffraction of the telescope pupil suprimposed on the disc H2 image. The bright circular feature on the east side (left) is arc-like and likely originates from a quasi-static speckle pinned on the residual of an Airy ring.}
        \label{fig_airy_rings}
\end{figure}

The final result was normalised to one at the NE ansa. The resulting curve for the H2 (top) and H3 filters (bottom) are shown in Fig. \ref{fig_phase_function}. No spectral dependance in the phase function is observed between the two filters, within error bars. The curves show a steep decrease from the smallest scattering angle $\varphi=90^\circ-i=13.6^\circ$ to $40^\circ$ followed by an increasing and linear trend until the largest scattering angle $\varphi=90^\circ+i=166.6^\circ$ seen given the disc viewing angle. The increase detected beyond $160^\circ$ on the north side is likely to be an artifact resulting from a quasi-static speckle pinned on an Airy ring at this exact location and smeared as an arc due to the derotation of the images. There are several reasons for which we do not believe in this feature. First it is not seen on the southern side of the disc, although the SPF as derived from the northern and southern side must be identical. This bright feature clearly appears in the final image as a portion of a circular ring whereas the disc curvature is very small along the semi-minor axis. Last, we suspect that it may correspond to the location of a PSF Airy ring for the H2 and H3 wavelengths, as shown in Fig. \ref{fig_airy_rings}.
The sharp increase of the phase function for scattering angles below $30^\circ$ is interpreted as forward scattering, meaning that the western side is inclined towards the Earth. Although this has been a matter of debate \cite[see for instance][]{Milli2015,Perrin2015}, the phase function analysis now clearly supports this assumption. Indeed \citet{Hapke2012} analysed 495 varieties of particles including solar system regolith samples, volcanic ash as well as various minerals and derived a general trend for scattering particles known as the hockey stick relation. This empirical relation shows that there are no particles with a narrow backward scattering peak, as it would be the case if we consider that the NW side is inclined away from us. By keeping the same naming conventions as in \citet{Hapke2012}, this statement corresponds to the fact that a large shape parameter $b$ and positive asymmetry parameter $c$ corresponds to an empty region in their Fig. 2 and is therefore highly unlikely. The author interprets this behaviour in terms of intrinsic particle structure: in order to backscatter light efficiently, particles need to have a high density of internal or external scattering structures but these would in turn scatter light in a wide and low peak.

\subsection{Spectral reflectance}

To extract the spectral reflectance of the dust, we measured the photometry of the disc over the spectral range covered by IRDIS and the IFS. Because the disc is not seen with the same S/N in all regions for IRDIS and the IFS, we used two different regions to measure the photometry. First we used the whole area of the disc. Second, we excluded the ansae, which for the IFS are partially (northern ansa) or totally (southern ansa) cropped by the field of view, and the regions very close to the star within 0.25\arcsec{} where a few brighter speckles still contaminate the fainter western semi-minor axis in the IFS (see Fig. \ref{fig_ifs_mcADI_binned}). This region therefore includes only the central region of disc, spanning from 0.3\arcsec{} to 0.9\arcsec{} and probes scattering angles from $17^\circ$ to $59^\circ$ and $121^\circ$ to $164^\circ$.

\begin{figure}
        \centering
        \includegraphics[width=\hsize]{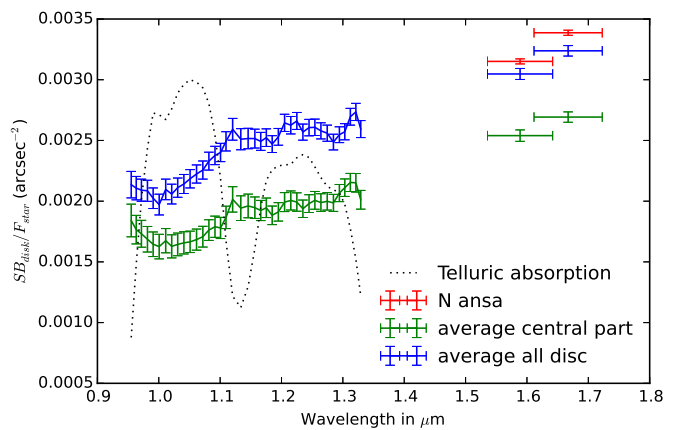}
        \caption{Spectral reflectance of the dust, averaged over the northern ansa (red points), the whole disc (blue points) or the central part of the ring between 0.3\arcsec{} and 0.9\arcsec{} (green points). The black dotted line corresponds to the scaled flux of the A0V star HR\,4796\,A to show the transmission of the atmosphere and instrument in the IFS wavelength range. The IFS spectral resolution is 50. The uncertainty is given at a $3\sigma$ level.}
        \label{fig_reflectance}
\end{figure}

We performed aperture photometry in non-overlapping circular apertures of constant radius 63mas, corresponding to 1.5 resolution elements at the longest wavelength of 1.66\micron{}, for an aperture area of $0.012\arcsec$. These apertures were placed along the best projected ellipse derived in Table \ref{tab_ring_morphological}. To convert this surface brighness in contrast per arcsec squared, we divided this encircled flux by the aperture surface area and normalised it by the stellar flux at each wavelength. This step removes the stellar colour so that only the scattering efficiency of the dust remains (averaged over the scattering angles seen by all apertures). The stellar flux was computed using the non-coronagraphic image of the star (e.g. the PSF) in a larger aperture of radius 0.43\arcsec{} and 0.3\arcsec{} for IRDIS and the IFS respectively. This radius turned out to yield the best S/N for the stellar flux  because the non-coronagraphic image is background limited beyond this radius.
In a second step, we corrected for the convolution by the instrumental PSF and for the ADI flux losses, as detailed in Section \ref{sec_phase_function}.

The result is shown in Fig. \ref{fig_reflectance} and combines the IFS and IRDIS data. It shows a red dust, with a 50\% increase in the spectral reflectance from 1\micron{} to 1.6\micron. The slope is consistent between the IFS and IRDIS wavelengths. The IFS spectra shows a few features, although not to a significant level. In particular all features seem to be correlated with the star spectra, shown in Fig. \ref{fig_reflectance} which can be regarded as an ideal telluric reference given that the spectral type of HR\,4796\,A is A0V. In particular, the change of slope around 1.15\micron{} in the dust spectrum is very likely due to the telluric water absorption, as well as the smooth decrease in the reflectance between 0.95\micron{} and 1\micron. 

\section{Discussion on the dust properties}
\label{sec_discussion_dust_prop}

\subsection{Constraints from the scattering phase function}

\begin{figure}
        \centering
        \includegraphics[width=\hsize]{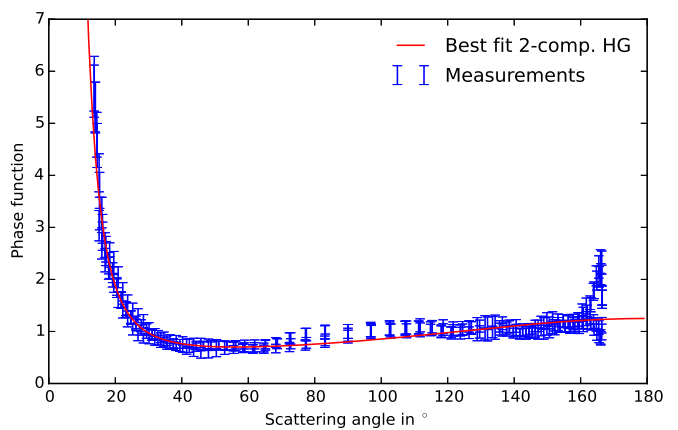}
        \caption{Fit of the averaged SPF with a 2-component HG function, parametrised by the two HG parameters $g_0=0.99^{+0.01}_{-0.38}$, $g_1=-0.14 \pm 0.006$ and the coefficients $a_0=4.0$, $a_1=0.82$. The uncertainty is shown at a $3\sigma$ level.} % error on a_1 \pm 0.012
        \label{fig_SPF_2comp_fit}
\end{figure}

The SPF of debris discs brings an insight in the properties of the scattering particles, as recently demonstrated using Saturn's G and D rings \citep{Hedman2015}. 
The most common way to describe the SPF is with the Henyey-Greenstein (HG) function \citep{Henyey1941}, which can be expressed as 
\begin{equation}
HG(g,\varphi)=\frac{1}{4\pi}\frac{1-g^2}{\left( 1-2g\cos{\varphi} +g^2 \right)^{3/2}}
.\end{equation}
Although this model has no physical motivation, it provides a useful way to describe the behaviour of the SPF because it depends on a single parameter $g$, known as the HG asymmetry parameter. It ranges between -1 for perfectly backward scattering and 1 for perfectly forward scattering, with 0 corresponding to isotropic scattering. The HG function is monotonic, so this already shows that it will not provide a good match to the measured SPF of HR\,4796\,A. However many authors \citep{Hapke2012,Stark2014, Hedman2015} show that multiple HG functions provide an adequate approximation to the measured SPF in planetary regoliths, planetary rings or dusty debris discs. We therefore fit the SPF with a 2-component HG function of the form
\begin{equation}
HG_2(g_1,w_1,g_2,w_2,\varphi) = w_1 HG(g_1,\varphi) +  w_2 HG(g_2,\varphi)
.\end{equation}
%The SPF used in the fit is the average between the measured SPF in the two filters H2 and H3 and between the two sides of the disc, re-interpolated at regular $\varphi$ intervals of 2 degrees between $15.5^\circ$ and $165^\circ$ to give a uniform weight to all scattering angles. Using the average between the Northern and Southern sides enables to smooth out any asymmetry due to local density enhancements in the ring not attributable to a change in the SPF. 
We combined the measurements from the north and south sides of the ring. The best fit yields a reduced chi squared of $1.08$ and is shown in Figure \ref{fig_SPF_2comp_fit}. The HG coefficents are $g_0=0.99^{+0.01}_{-0.38}$ for the forward scattering component, with a weight of 83\%, and $g_1=-0.14 \pm 0.006$ for the backward component, with a weight of 17\%. This function explains well the overall shape, both the steep forward scattering regime, the flat region around $50^\circ$ and the positive slope beyond $50^\circ$. Two regimes at about $100^\circ$ and $145^\circ$ are not well explained because the SPF has oscillations that cannot be captured by this simple model. The very high HG coefficient for the foward scattering component $g_0$ indicates that the grains are overall much larger than the wavelength $\lambda=1.65\micron$. The HG coefficient of the backward scattering side $g_1=-0.14 \pm 0.006$ is compatible with the previous HST~/~NICMOS measurements from \citet{Debes2008} and was previously interpreted as forward scattering.

\begin{figure}
        \centering
        \includegraphics[width=\hsize]{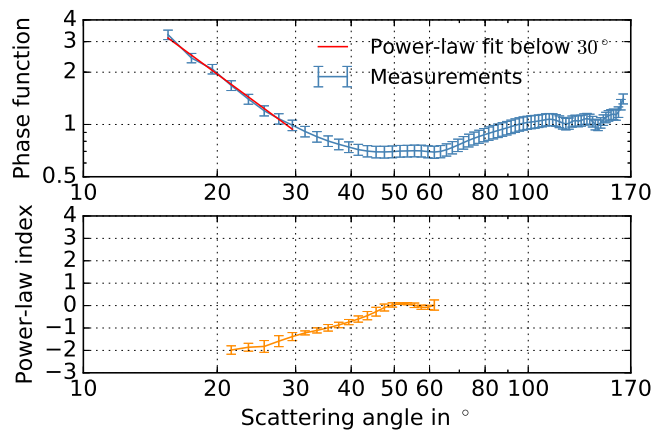}
        \caption{Top: Log-log plot of the averaged SPF with a power-law fit below $30^\circ$ (power-law index of $-1.88\pm0.06$). Bottom: Power-law index as a function of the scattering angle, for a fit performed on a sliding range of scattering angles of length $14^\circ$. This index can be related to the dust size distribution for small scattering angles.}
        \label{fig_SPF_power_law}
\end{figure}

The Fraunhofer diffraction theory is the simplest theory that can be used to relate the SPF with information on the properties of the scattering particles.  It is valid for opaque particles of size $s$ at scattering angles small with respect to $\lambda/s$. The grains of the HR\,4796 disc are thought to be aggregates of elementary size about 1\micron{} \citep{Milli2015}, meaning that the approximation of the SPF by diffracting particles is valid for $\varphi \ll 92^\circ$. In this limit, if the differential size distribution is a pure power law $dN/ds \propto s^\nu$, the resulting SPF would be a power-law function of scattering angle proportionnal to $s^{-(\nu+5)}$ \citep{Hedman2015}. A log-log plot of the SPF (Fig. \ref{fig_SPF_power_law}, top) shows that the values corresponding to the smallest accessible scattering angles below $30^\circ$ are well fit by a power-law function of index $-1.88\pm0.06$ (red curve). We also show in the bottom panel the local value of the power-law index in the SPF as a function of scattering angle. To do so, we fit the SPF in the vicinity of each data point (within $\pm7^\circ$) to a simple power-law function SPF $\propto \varphi^r$ where $r$ is the power-law index.

Assuming that we already comply with the Fraunhofer validity limit at $\varphi \sim 20^\circ$, we notice that the power-law index converges to $-2 \pm 0.1$, which translates into a size distribution $dN/ds$ proportional to $s^{-(-2+5)}=s^{-3}$ \citep{Hedman2015}. This result is in agreement with previous modelling suggesting that the size distribution $dN/ds$ is likely following the traditional $s^{-3.5}$ \citep{Milli2015} expected for a collisional cascade in steady state \citep{Dohnanyi1969}. However this result must be taken with caution, because the power-law index fit to the SPF is not exactly constant for small scattering angles (see Fig. \ref{fig_SPF_power_law} bottom), suggesting that either we are no longer in the approximation of Fraunhofer diffraction at $\varphi=21.5^\circ$ or that the size distribution does not follow a perfect power-law.

\begin{table}
\caption{Grid of parameters for the 15600 models generated.}             % title of Table
\label{tab_grid}      % is used to refer this table in the text
\centering                          % used for centering table
\begin{tabular}{p{1.7cm} p{1cm} p{1cm} p{1cm} p{1.3cm}}        % centered columns (4 columns)
\hline\hline                 % inserts double horizontal lines
Parameter & Min. value & Max. value & $N_\text{sample}$ & Sampling  \\    % table heading 
\hline    
Scattering theory &  \multicolumn{2}{c}{Mie / DHS} & / & / \\   % inserts single horizontal line
 $s_{min}$ (\micron) & 0.1 & 100 & 13 & log. \\      % inserting body of the table
$p_{H_2O}$ (\%) & 1 & 90 & 5 & log. \\      % inserting body of the table
$P$ (\%) & 0 & 80 & 5 & linear \\      % inserting body of the table
$q_\text{Sior}$ & 0 & 1 & 6 & linear \\
$\nu$ & -2.5 & -5.5 & 4 & linear \\
\hline                                   %inserts single line
\end{tabular}
\end{table}

\begin{figure}
        \centering
        \includegraphics[width=\hsize]{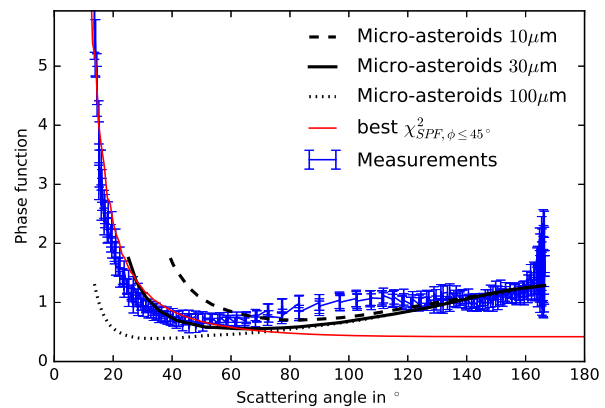}
        \caption{Comparison of the measured SPF (blue bars) with the theoretical SPF predicted by the Mie/DHS theory for a dust composition and size distribution best matching the observations (red curve). The three black curves show the SPF predicted by the Hapke theory with the assumptions of regolith particles for three different grain sizes from \citet{Min2010}, described as micro-asteroids by these authors. The uncertainty is given at a $3\sigma$ level.}
        \label{fig_fit_Mie_DHS}
\end{figure}

The Mie theory extends the results of the Fraunhofer diffraction theory by providing the exact analytical expression of the SPF over all scattering angles under the assumption that the scattering particles are spherical dielectric grains. In \citet{Milli2015}, we computed the theoretical SPF for a sample of 7800 dust compositions and sizes, using the Mie theory or the Distribution of Hollow Spheres \citep[DHS,][]{Min2005}. We therefore used the same models to investigate the compatibility of our new measurements. As a reminder, these models are based on a porous dust grain composed of a mixture of astronomical amorphous silicates, carbonaceous refractory material and water ice partially filling the holes created by porosity. We kept as much as possible the same notations as in \citet{Augereau1999} namely a porosity without ice $P$, a fraction of vacuum removed by the ice $p_{H_2O}$, a silicate over organic refractory volume fraction $q_\text{Sior}$. The minimum grain size was renamed $s_{min}$ to avoid confusion with the semi-major axis $a$, and the  index of the power-law size distribution was renamed $\nu$.  The grid of these five free parameters is described in Table \ref{tab_grid}. In \citet{Milli2015}, the SPF was measured directly on the synthetic disc image for scattering angles between $75^\circ$ and $105^\circ$ only. Only DHS models with large $10\micron$ silicate grains could reproduce the local brightness enhancement of the backward side close to the ansa, but the models predicted a strong forward-scattering peak whose reality could not be tested.
%\textbf{For the DHS models, we used a denser sampling of the fraction $f$ occupied by the central vacuum inclusion \citep[see][for details]{Min2005}, to remove unphysical oscillations observed in the predicted SPF related to undersampling}
Despite this exhaustive grid over a wide range of grain sizes, distribution and composition, none of the SPF predicted by the Mie or DHS theory is compatible with the measured SPF of HR\,4796\,A between $\varphi=13.6^\circ$ and $166.4^\circ$. The smallest reduced chi squared is 131. None of the models can simultaneously explain the forward scattering peak below $30^\circ$ and the increase of the SPF beyond $60^\circ$. As a result, the best chi squared models are always a trade-off between these two regimes, but are never able to well reproduce any of them. We therefore tried to find among our grid of models those providing a good fit to the forward scattering part of the PSF only, for $\varphi \leq 45^\circ$. The best model is shown as a red curve in Fig. \ref{fig_fit_Mie_DHS}, where the blue bars are our measurements. The reduced chi squared is 3.8. The predicted SPF is mostly sensistive to the minimum grain size and size distribution of the dust population. Interestingly the best model for $\varphi \leq 45^\circ$ is obtained with a very steep distribution of power-law $\nu=-5.5$ with a minimum grain size $s_{min}=17.8\micron$. This means that the dust population is strongly dominated in scattered light by the $17.8\micron$ particles. This size is much larger than the wavelength $\lambda=1.66\micron$, we are therefore in the regime of geometric optics and this finding corroborates the very high HG coefficient of $0.99$ derived for this range of scattering angles. We note that our grid for the parameter $s_{min}$ is logarithmic and therefore not very well sampled around 20\micron : it steps from $10$, to $17.8$ and $30\micron$, therefore a better match of the SPF for  $\varphi \leq 45^\circ$ is expected with a finer grid. We note that this slope of -5.5 is incompatible with the Fraunhofer  model which favoured a slope of -3, indicating the limits of this simple model at $\varphi$ as large as $20^\circ$ or of our assumption of a power-law size distribution of spherical particles.

This value can interestingly be compared to the blowout size of the grains, which is defined as the maximum size below which grains are placed on an unbound orbit. For such a critical size, the ratio between radiation pressure to gravity equals one half. \citet{Augereau1999} derived a blowout size of 10\micron{} from the spectral energy distribution (SED) modelling, which is the same order of magnitude as the $17.8\micron$ value derived from the SPF analysis. However, this theoretical blowout size assumes that the grains are spherical and homogeneous, but the effect of radiation pressure on inhomogeneous aggregates can induce a significant departure to this value, as shown by \citet{Saija2003}, and can very well explain the discrepency seen here. 

\subsection{Constraints from the spectral reflectance}

\begin{figure}
        \centering
        \includegraphics[width=\hsize]{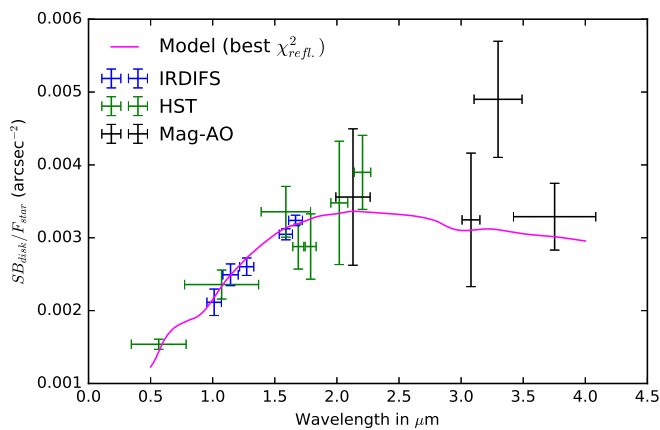}
        \caption{Reported values of the spectral reflectance of the dust from \citet{Rodigas2015} with the new data from Fig. \ref{fig_reflectance} (average value over the disc) that used a circular aperture of radius 63mas. The graph also shows the best model explaining the data described in Table \ref{tab_chi2_Mie_DHS}. The uncertainty is given at a $3\sigma$ level}
        \label{fig_reflectance_HST}
\end{figure}

The result agrees with previous values of the spectral reflectance as measured through broadband photometry with HST / NICMOS \citep{Debes2008}, HST / STIS \citep{Schneider2009} and MagAO Clio-2 \citep{Rodigas2015}. We summarised previous measurements in Fig. \ref{fig_reflectance_HST}.

We used the models presented in \citet{Milli2015} and in Section \ref{sec_dust_prop} to identify the best matching models in that grid. We used the measured reflectance averaged over the whole disc, because there is no significant difference with the measurements corresponding to the central part only and compared it to the predicted values. These predicted values were computed by averaging the Stokes parameter $S_{11}(\theta)$ sampled at the same scattering phase angles as those corresponding to the apertures used for the photometric measurements. We did not perform an absolute flux comparison because the mass of dust grains is not well constrained\footnote{For instance, \citet{Augereau1999} estimated 3.9\Mearth{} of dust between 10\micron{} and 1.4m for their best fit model to the SED, while \citet{Milli2015} derived a mass of 0.5\Mearth{} of dust between 1\micron{} and 1cm}, but we scaled the predicted reflectance to match that of the disc. To avoid overweigthing the wavelength range corresponding to the IFS which has 39 measurement points, we binned the IFS measurements in three spectral channels only.

Many models give acceptable solutions with a chi squared of about one, however none of them are also compatible with the disc SED or the phase function. The goodness of fit of each model with the disc SED was computed in \citet{Milli2015}. It used eight SED measurements above 20\micron. The best model matching the spectral reflectance has a composition similar to the best model matching the SED but differs in terms of grain minimum size. It uses small particles of minimum size $0.3\micron$ smaller than the blowout size, whereas the best SED model uses a minimum grain size of 1.78\micron. We will now compare this observable to the SPF and SED.

\subsection{Bayesian analysis}

\begin{figure*}
        \centering
        \includegraphics[width=\hsize]{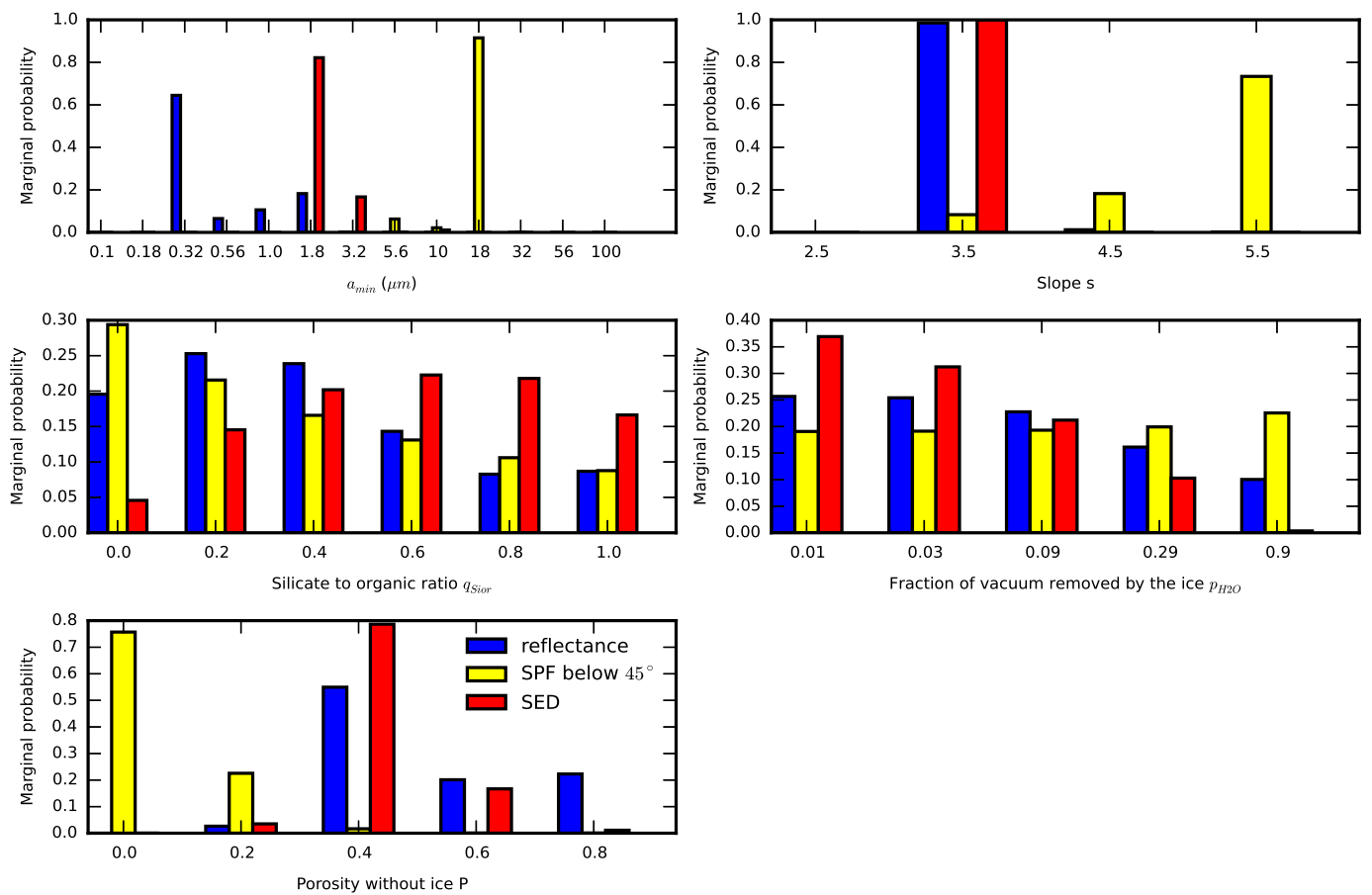}
        \caption{Marginal probability distributions of the five free parameters of our model, based on fitting the SED (red bars), the spectral reflectance of the disc (blue bars), or the SPF below $45^\circ$ (yellow bars). These distributions are derived from models created using the Mie theory.}
        \label{fig_marginal_proba_DHS}
\end{figure*}

In Section  \ref{sec_dust_prop} we provided two new independent measurements of the dust properties: the SPF and the spectral reflectance. To study the constraints they bring on the dust population, we performed a Bayesian analysis as it was done in \citet{Milli2015}. Each model was assigned a probability that the data are drawn from the model parameters. We do not have any a priori information on these parameters, we therefore assumed a uniform prior, corresponding to a uniform sampling of our parameters by our grid. Under this uniform prior assumption, the probability $\Psi$ that the data corresponds to a given parameter set is given by
\begin{equation}
\Psi=\Psi_0\text{exp}\left( -\frac{\chi^2}{2}\right)
,\end{equation}where 
$\Psi_0$ is a normalisation constant introduced so that the sum of probabilities over all models is unity. The probability given here is only valid within the framework of our modelling and parameter space.

Figure \ref{fig_marginal_proba_DHS} shows the inferred probability distribution for each of our five free parameters, after marginalisation against all other four parameters. It is shown here using the Mie theory. We compare the marginal probability derived from the SPF, from the reflectance spectra, and as a comparison point from the SED \cite[same data as presented in Fig. A.1 of][]{Milli2015}. Table  \ref{tab_chi2_Mie_DHS} summarizes the best chi-squared values and parameters for each observable.
% because the best scattered light models and overall models are obtained with this theory, but we provide in appendix \ref{App_marginal_PDF_DHS} the probability distributions obtained using the DHS theory.
Three parameters are significantly constrained by the available observables: the minimum grain size $s_\text{min}$, the slope of the size distribution $\nu$ and the porosity without ice $P$. The highest probable minimum grain size is not compatible between our three observables: while the SED favours grains of about 1.8\micron, the SPF indicates much larger particles of 18\micron, with a high probability. Such large Mie grains located 77 au from the star are much too cold to be consistent with the IRAS colours of the system. On the other hand, the reflectance favours small submicronic grains. Both the reflectance and SED agree well with a distribution size with slope $-3.5$ as theoretically predicted \citep{Dohnanyi1969}. However the SPF indicates a very steep distribution, hence dominated by the smallest particles of size 18\micron. 
 This mismatch between the predicted scattered light emission and thermal emission is now a common issue reported in debris disc modelling \cite[e.g.][]{Schneider2006,Rodigas2014,Milli2015,Olofsson2016}. Finding an overall model that fits all observed measurements is difficult, mainly because the spherical assumption made on the shape of the particles to deduce the scattered light properties is  not valid, as explained below. 

The most striking finding among our new observations concerns the SPF of the dust around HR\,4796\,A. The rise in the SPF beyond $60^\circ$ is reminiscent of the case of Fomalhaut, hosting a debris disc inclined by about $65^\circ$. In the Fomalhaut case, the smaller inclination of the system does not permit us to detect any peak of forward scattering because the observations only probe scattering phase angles above $25^\circ$. However spectrally resolved interferometric observations \citep{LeBouquin2009} derived the orientation vector of the stellar rotation axis and showed that the brighter side of the system is inclined away from the Earth, assuming that the disc is in the stellar equatorial plane. The best interpretation so far implies very large dust grains of 100\micron{} whose diffraction peak is confined within a narrow range of scattering angles undetected from the Earth given the system inclination \citep{Min2012}. In this scenario, the backward scattering nature of the grains is explained by the fact we see the phases of the grains, as in the cases of the Moon or asteroids. To sustain this scenario, \citet{Min2012} assumed the dust SPF behaves like regolith particles described by the Hapke theory, and included the contribution of diffraction. The result shown in Fig. 2 of \citet{Min2012} and reproduced for our purpose in our Fig. \ref{fig_fit_Mie_DHS} (referred to as micro-asteroids in the caption) is very similar to our measured SPF. It displays a diffraction peak at small scattering angles followed by an increasing SPF at greater angles. The transition between the two regimes depends on the size of the particles: the larger the particles, the narrower the diffraction peak. Figure \ref{fig_fit_Mie_DHS} reproduces their curves for particles of 10\micron, 30\micron{} and 100\micron. Our measurements are compatible with their model for a particle size between 10 and 30\micron, closer to their 30\micron{} model. This is very consistent with the fact that our best match of the SPF by our grid of models correspond to a dust population of particle minimum size about 20\micron. 

Another piece of evidence suggesting the presence of large aggregates comes from the findings of \citet{Min2016} who showed that aggregates show a flat or increasing SPF when modelled assuming complex irregular shapes particles, whereas all simplified methods such as Mie or DHS predict a decreasing phase function with scattering angle. In fact, they show that large aggregate particles display a phase function with a strong forward-scattering peak and mildly backward-scattering part of the phase function. This backward-scattering part of the phase function is not found for smooth particles, and this region grows with the aggregate size. Our measurement matches qualtitatively very well this theoretical prediction if the dust particles of HR\,4796\, A are aggregates of size about 20\micron. A quantitative analysis using advanced models able to predict the SPF for irregular inhomogeneous particles such as the discrete dipole approximation \cite[DDA,][]{Purcell1973} is beyond the scope of this paper. 
One major problem remains, which is that these large particles would emit much too efficiently in the far-infrared, thus would be cold. The predictions for the emitted 25\micron{} flux density of the ring are an order of magnitude too small when these large grains are used, hence the discrepency seen in Fig. \ref{fig_marginal_proba_DHS} between the more likely minimum grain size deduced from the SPF and the SED or  the spectral reflectance. Such a discrepency can arise in the case of aggregates if different observables are sensitive to the different scales of the particle. It was shown indeed that for sufficiently fluffy grains, the SPF behaviour is driven by the overall size of the aggregate, whereas the degree of polarization is expected to be determined by the size of the constituent particles \citep{Min2016,Olofsson2016}. Similarly, analytical studies of aggregates using the effective medium theory (EMT) have shown that the absorption mass opacity of fluffy aggregates is characterised by the product $a \times f$ , where $a$ is the dust radius and $f$ is the filling factor \citep{Kataoka2014}. This means that the opacity is almost independent of the aggregate size when $a \times f$ is constant. We could, therefore, obtain the same opacity between more compact $\sim2\micron$ grains and larger 20\micron{} aggregates if the filling factor is smaller by a factor 10, thus reconciling the multimodal probability distribution seen in Fig. \ref{fig_marginal_proba_DHS} (first panel). There are however two cautionary remarks. First, the EMT is  known to be accurate for porous aggregates whose constituent particles are small compared to the wavelength of incident radiation, which is only marginally validated for HR\,4796 if the monomeres are $\sim1-2\micron{}$. Second, porous aggregates are thought to form in protoplanetary discs through slow sticky collisions in a high density medium but are more difficult to make in debris discs where collisions are destructive, with higher velocities, in a low density medium.

The strong forward-scattering peak detected in these new observations also sheds a new light on the polarised image of the disc presented in \citet{Perrin2015} and \citet{Milli2015}. The SPF is indeed six times higher on the semi-minor axis at $\varphi=13.6^\circ$ (NW side) than in the ansae of the ring. In polarised light, the NW side is also the brightest part of the disc. This has triggered many questions because the polarised fraction is not expected to peak at such a small scattering angle. 
The fact that the NW semi-minor axis now also appears as the brightest part of the ring in unpolarised light shows that the polarised fraction drops by a factor smaller than six between $\varphi=13.6^\circ$ and $\varphi=90^\circ$. Future observations combining polarised and unpolarised light are clearly needed to measure the degree of polarisation at small scattering angles, and therefore constrain the size of the aggregate elemental constituents. Such observations could validate the presence of aggregates by comparing the shape of the unpolarised and polarised SPF. Such a study was undertaken for HD\,61005 using VLT / SPHERE  \citep{Olofsson2016}: a similar very narrow peak of forward scattering is detected in unpolarised light, but the higher inclination of the system makes retrieving and modelling the SPF more difficult. The dependance of the SPF with wavelength represents another property to investigate in future observations to understand the discrepency in the particle size. For instance, in the Mie theory for particles of the same order of magnitude as the wavelength, a flip in the sign of the polarisation fraction is expected between the forward and backward scattering side. Polarimetric observations from the optical to the near-infrared can test this prediction. Lastly, resolved submillimetre observations will permit us to validate our assumption of a uniform azimuthal dust density distribution and refine the dust mass estimation. This estimation is critical to validate that the disc is optically thin and perform an absolute comparison of the spectral reflectance with that predicted by models. Combining resolved scattered light and submillimetre observations will also permit us to break the degeneracies between the radial and vertical dust density distribution. The disc vertical extent is often used as a measure of the dynamical stirring to trace the presence of hidden perturbers such as planetary embryos  \citep{Thebault2009}, which might also explain the sharp offset ring.

%Future observations at different wavelengths to measure the width of the disc at various wavelength and see if it is wider to test the otpical thickness of the ring.
%Observe it with the IFS in extended mode to orbe the water aborption band.
%Polarized observations to retrieve the polarized fraction over all the disc.

\begin{table}
%\begin{sidewaystable*}
\caption{Goodness of fit estimates and corresponding parameters for the best models with respect to the SED or the scattered light observables.}            
\centering          
\label{tab_chi2_Mie_DHS}      
\begin{tabular}{ c | c c c }     % 7 columns 
\hline\hline       
  & best & best    & best  \\
  & SED\tablefootmark{a} & reflectance & SPF $\varphi \leq 45^\circ$ \\
  \hline                  
Theory               & DHS    & Mie & Mie \\
$\nu$                 & -3.5     & -3.5 & -5.5  \\
$q_\text{Sior}$  & 0.2       & 0.2 & 0.0 \\
$p_{H_2O}$      & 3.1\%   & 1.0\% & 90\% \\
$s_\text{min}$  & 1.78      & 0.3 & 17.8 \\
P                       & 20.0\%      & 40.0\% & 0.10\% \\
\hline                  
$\chi^2_\text{SED}$                     & \textbf{1.7} & 129                & 48          \\
$\chi^2_\text{refl.}$                      & 352             & \textbf{1.4}  & 6.2           \\
$\chi^2_\text{SPF}$                     & 394             & 442               & \textbf{3.8} \\
\hline
\end{tabular}
\tablefoot{
\tablefoottext{a}{Best model explaining the SED already presented in \citet{Milli2015} and displayed here as a reference.}
}
\end{table}
%\end{sidewaystable*}

\section{On the origins of a sharp offset ring}
\label{sec_origin_offset_sharp}

To be maintained over the age of the system, the sharpness and eccentricity of the ring require a confinement mechanism. The new constraints presented here enable us to re-examine the past scenarios proposed to explain this peculiar morphology.

\subsection{The influence of a stellar companion}

The ring can be maintained eccentric by an inner (sub)stellar companion orbiting HR\,4796\,A. In this scenario, the disc is circumbinary and its centre is the centre of mass of the binary system. Past observations combining sparse aperture masking with direct imaging at \Lp{} carried out with VLT/NaCo can already rule out this scenario \citep{Lagrange2012}.

\begin{figure}
        \centering
        \includegraphics[width=\hsize]{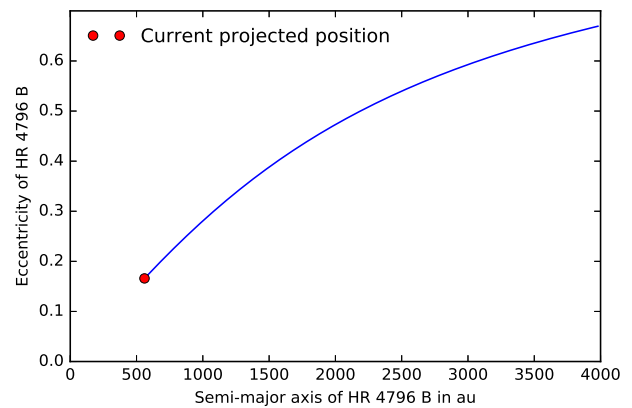}
        \caption{Relation between the semi-major axis of HR\,4796\,B and its eccentricity if it is responsible for the offset of the ring through the secular perturbation mechanism.}
        \label{fig_eccentricity}
\end{figure}

The presence of the binary companion HR\,4796\,B at a projected distance of about 560\,au, can also explain the disc outer truncation and its eccentricity. Indeed an eccentric perturber can place the ring in an eccentric orbit through secular interaction \citep{Augereau2004,Wyatt2005}. When secularly perturbed, the forced eccentricity $e_f$ of a planetesimal evolves cyclically to reach a maximum value $e_{f,max}$ which can be related to the eccentricity of the perturber, $e_p$, through Eq. 2. of \citet{Faramaz2014} (second order approximation in $a/a_p$):
\begin{equation}
e_{f,max} = \frac{5}{2}\frac{a e_p}{a_p (1-e_p^2)}
,\end{equation}
where $a$ and $a_p$ are respectively the semi-major axis of the disc and perturber. From this relation and by using the projected separation of HR\,4796\,B of 560\,au as a lower bound for the perturber semi-major axis $a_p$,  one can derive that the minimum perturber eccentricity $e_p$ necessary to produce the measured disc eccentricity of 0.06 is 0.17. It is represented as the red dot in Fig. \ref{fig_eccentricity}. For a larger semi-major axis, a higher eccentricity is required to drive the secular perturbation mechanism, as shown by the blue curve.

\subsection{Planets sheperding the ring}

 Alternatively, the eccentricity can be forced by the secular gravitational interaction with a planet on a similar eccentric orbit interior to the ring \citep{Wisdom1980}, which is analysed in this subsection. This scenario can also explain why the inner edge is so steep. It is, however, not sufficient to explain why the outer edge is also very steep. This latter feature might be due to the presence of an unseen planetary companion beyond the outer edge of the ring, although \citet{Thebault2012} showed that planets are less effective than stellar companions to truncate the outer edge of a debris disc. The steepness of the surface brightness profile is difficult to compare to simulations as planet-less scenarios can also display steep profiles beyond the outer edge over a limited radial distance \citep{Thebault2012}. 

\begin{figure}
        \centering
        \includegraphics[width=\hsize]{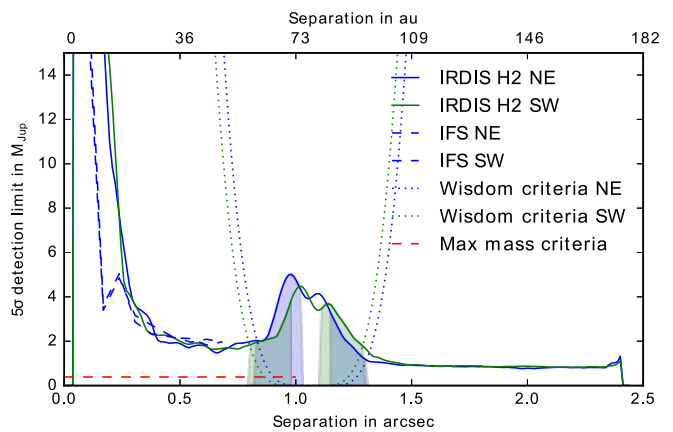}
        \caption{Detection limits in Jupiter mass as a function of the distance to the star along the semi-major axis of the disc. The blue curves correspond to the NE ansa and the green curve to the SW ansa.The dotted lines show the mass distance relationship of a planet shaping the edge of the disc. The shaded area represents the region in the ($M_\text{pl}$,$a_\text{pl}$) parameter space compatible with a planet shepherding the disc, given the sensitivity of our new measurements along the semi-major axis. It is defined as the region where the sensitivity is below the critical mass for the Wisdom criteria, excluding the area of the disc.}
        \label{fig_mass_limits_NE_SW}
\end{figure}

\begin{figure}
        \centering
        \includegraphics[width=\hsize]{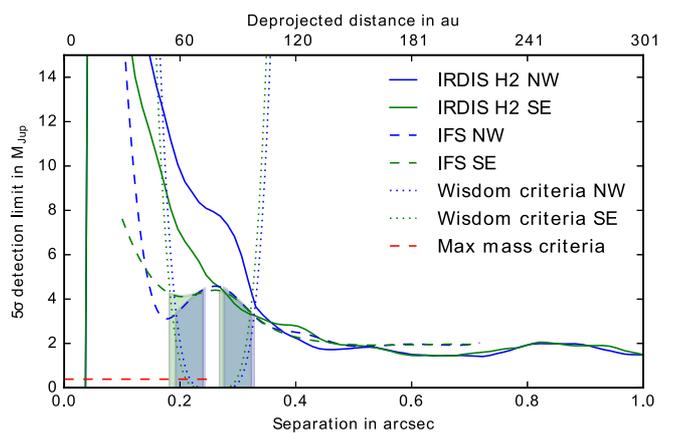}
        \caption{Detection limits in Jupiter mass as a function of the distance to the star along the semi-minor axis of the disc. The blue curves correspond to the NW ansa and the green curve to the SE ansa.The dotted lines show the mass distance relationship of a planet shaping the edge of the disc. The shaded area represents the region in the ($M_\text{pl}$,$a_\text{pl}$) parameter space compatible with a planet shepherding the disc, given the sensitivity of our new measurements along the semi-minor axis.}
        \label{fig_mass_limits_NW_SE}
\end{figure}

Therefore, we focus our analysis on constraining the mass of a planet shaping the inner edge of the disc. Several authors provide guidelines to relate the mass of a shepherding planet with a ring observable parameter. Analytical considerations show that the width $\delta a$ of the chaotic zone created by overlapping mean-motion resonances of a planet on a circular orbit is given by 
\begin{equation}
\label{eq_Wisdom}
\delta a/a=1.3\mu^{2/7}
,\end{equation}
where $\mu$ is the ratio of the planetary to stellar mass and a is the planet semi-major axis. \citet{Mustill2012} extended this result with numerical simulations to the case of a planet on an eccentric orbit, showing that Eq.\ref{eq_Wisdom}  still applies when the eccentricity is kept below the critical value of $0.21\mu^{3/7}$. In the case of HR\,4796\,A, assuming the planet eccentricity drives the disc eccentricity and is therefore also $0.058$, the critical mass under which Eq. \ref{eq_Wisdom} remains valid is $125M_{\text{Jup}}$ which is much larger than our upper limits. We can therefore apply Eq. \ref{eq_Wisdom} and combine it to our sensitivity to point source objects in order to exclude incompatible combinations in the ($M_\text{pl}$,$a_\text{pl}$) parameter space, as already done in \citet{Lagrange2012}. The result is shown in Fig. \ref{fig_mass_limits_NE_SW}, for the semi-major axis and in Fig. \ref{fig_mass_limits_NW_SE} for the semi-minor axis of the disc and is labelled as Wisdom criteria in the legend. Because the distances along the semi-minor axis of the disc need to be deprojected to compare them to the semi-major axis, the constraints are tighter along the semi-major axis. We conclude from this analysis that a planet with a mass and distance to the ring within the shaded blue and green area can be responsible for the observed morphology of the ring. 

In order to provide observers with a simple way to connect the properties of a disc-shepherding planet with disc observables, \citet{Rodigas2014} derived a linear expression relating a shepherding planet's maximum mass to the debris ring's observed width in scattered light. This relation was derived performing dynamical N-body simulations and assumes a single planet orbiting a single star interior to a debris disc, coplanar with the planet. They showed that the normalised full width at half maximum nFWHM of the disc scales linearly with the maximum mass of the shepherding planet, independently of the initial eccentricites of the planetesimal belt or of the integration time
\begin{equation}
m_p/M_J = \left( \frac{nFWHM-(0.107 \pm 0.032)}{0.019 \pm 0.0064} \right) \left( \frac{M_{\ast}}{M_\odot} \right)
.\end{equation}
From the measurements detailed in Table \ref{tab_radial_prof}, we find the averaged nFWHM to be $11\% \pm  7\%$ (based on a FWHM of $0.117\arcsec \pm  0.070\arcsec$ average of the non-ADI images in H2 and H3 for the NE and SW ansae), which yields a shepherding planet maximal mass of $0.4 M_J \substack{+8.4 \\ -0.4}$, or 1.3 Saturn mass. In Figs. \ref{fig_mass_limits_NE_SW} and \ref{fig_mass_limits_NW_SE} we compare this threshold to our sensitivity along the semi-major and minor axis of the ring. It is much tighter than our sensitivity, as we are not able to exclude the presence of a planet below 2\Mjup interior to ther ring, but the large upper uncertainty on this threshold weakens this results.

We therefore conclude that the narrow width of the ring points towards a Saturn mass planet or lighter, which we cannot rule out given our sensitivity. If it orbits interior to the ring and is shaping the edge, the Wisdom criteria shows that such a planet must be within 6 au from the ring inner edge. Any planet that is less massive and closer to the inner edge would also pass unnoticed.

\subsection{Gas as the origin of the sharp ring}

%Last, the presence of gas can also confine the dust in a narrow offset ring. A recent study has shown that the observed eccentric ring morphology in several debris discs can be caused by the interactions with gas \citep{Lyra2013}, through a photoelectric instability caused by the feedback of dust on gas through photoelectric heating. For this instability to take place, gas to dust ratios equal to or greater than one are required. This hypothesis is particularly relevant given the recent discoveries of gas in several debris discs \cite[e.g.][]{Moor2015}. Two origins for this gas have been proposed: a secondary  origin, where volatiles are continuously released from ices on planetesimals \citep{Dent2014, Matra2017} and a primordial origin, which requires some delay in the dispersal of gas from the protoplanetary disc \citep{Kospal2013}. 

In the primordial, protoplanetary disc-like scenario, the gas is mainly molecular, and is dominated by CO and H$_2$. Given the high H$_2$ densities expected in such an environment, CO is in local thermodynamic equilibrium (LTE), meaning that the observed CO flux depends on the total CO mass and on the gas kinetic temperature $T_k$ \cite[Eq. 2 and 9 in ][]{Matra2015}. The object HR\,4796\,A has been searched for CO several times at millimetre wavelengths with no success, through its J=1-0 \citep{Liseau1999}, J=2-1 \citep{Zuckerman1995}
and J=3-2 \citep{Hales2014} transitions. Upper limits (3$\sigma$) achieved were 0.91, 0.64, and 2.03 Jy km/s. Varying temperatures between 20 K (freeze-out temperature of CO) and 200 K, and assuming optically thin emission, the total CO mass can be constrained to be within $0.8-1.4 \times 10^{-4}$ M$_{\oplus}$. Assuming a CO to H$_2$ relative abundance of 10$^{-4}$ (typical of the ISM and largely assumed for protoplanetary discs), we can then set an upper limit on the total gas mass of 0.06-0.11 M$_{\oplus}$.

In the secondary scenario, H$_2$ is absent and the gas is composed of CO and other molecules that may be released from the comets, as well as their photodissociation products. A recently developed thermodynamical model following the evolution of exocometary gas \citep{Kral2016, Kral2017subm} shows that these photodissociation products are likely to dominate the gas mass, since molecules including CO are photodestroyed very rapidly after release (in less than 100 years) due to the strong UV radiation environment found around an A0V star. The same radiation field is able to ionise these atomic species, explaining the detection of CII around several debris discs \cite[e.g.][]{Riviere2014}. HR\,4796\,A was also searched for atomic oxygen (OI) and ionised carbon (CII) emission by \textit{Herschel} \citep{Meeus2012}, which set upper limits of 6.3 and 5.4 $\times 10^{-18}$ Wm$^{-2}$, respectively. Due to the low gas densities expected in this scenario, species with high critical densities such as CO and OI are expected to be out of LTE, meaning that fluxes are dependent on the density of collider species and on the radiation field as well as on the gas temperature and OI mass \cite[Eq. 7 in][]{Matra2015}. This means that the derivation of OI mass upper limits from the observed fluxes is highly degenerate; the \textit{Herschel} observations therefore cannot tightly constrain the oxygen mass in these low-density discs. On the other hand, taking electrons as the dominant collider species \cite[as shown to be likely in secondary gas discs,][]{Matra2015}, CII has a much lower critical density of order 5-10 electrons cm$^{-3}$ \citep{Goldsmith2012}. The electron density predicted by this exocometary model in HR\,4796\,A is much higher \cite[of order $5\times10^{4}$ at the location of the belt,][]{Kral2017subm}. This means that LTE is a good approximation for CII, yielding a mass upper limit between (0.1 - 2.5) $\times10^{-2}$ M$_{\oplus}$ for the same 20-100 K temperature range considered above for CO. However, our secondary gas model predicts an average ionised fraction in the disc that is very low (M$_{\rm CII}/M_{\rm CI}\sim0.1$), meaning that we expect a much higher total carbon mass in HR\,4796\,A ring of $\sim$0.19 M$_{\oplus}$. Then, assuming the dominant molecule released by the comets to be CO, yielding a C/O abundance of one, we can estimate the total gas mass in the secondary origin scenario to be 0.44 M$_{\oplus}$.
However, we note that the \citet{Kral2016} model shows that the bulk of the gas will rapidly expand viscously to form an accretion disc. This means that only $\sim$6\% of this total gas mass (0.03 M$_{\oplus}$) will be co-located within the $7.2$ AU wide dust ring.

In summary, the secondary model predicts a gas mass co-located with the ring of 0.03 M$_{\oplus}$, whereas in the primordial scenario we expect a total gas mass of at most 0.11 M$_{\oplus}$. On the other hand, SED modelling of the ring's infrared excess \citep{Milli2015} yields a total dust mass in the ring of 0.5 M$_{\oplus}$ distributed among particles between 1\micron{} and 10 mm. Regardless of the gas origin scenario, we therefore derive an upper limit on the gas to dust ratio at the ring's location of $\lesssim0.2$, which rules out the presence of gas in sufficient amounts to explain the observed ring morphology in the HR\,4796\,A system. A similar conclusion was recently achieved for the narrow HD\,181327 debris ring \citep{Marino2016}, where very low levels of second-generation CO were discovered, but in amounts that are insufficient to create the ring morphology. Similarly sensitive ALMA observations of the HR\,4796\,A system (Kennedy et al. in prep) will be necessary to look for any secondary CO that may be released due to the high collisional activity within the ring, and to study the ice composition of the grains, which may in turn provide additional constraints on the grain properties inferred through our modelling in Sect. \ref{sec_discussion_dust_prop}.

\section{Conclusions}
\label{sec_conclusions}

Our new observations of the dust ring around HR\,4796\,A reveal for the first time the semi-minor axis of the disc in unpolarised light. It provides stringent constraints on the dust properties through the measurement of the scattering phase function and spectral reflectance, and on the presence of planets through the exquisite detection limits of the IFS at very short separation below 0.4\arcsec{} combined with IRDIS at larger separations. We have compared our measured scattered light properties to those of models derived  using the Mie and DHS theory and discussed the general properties of the dust particles able to reproduce such behaviours. We come to the following conclusions:

We measure with high accuracy the morphological parameters of the ring, and confirm the ellipticity of 0.06. We also measure and confirm the steep slopes of the brightness distribution along the ansae.

We show that the NW side of the disc appeared much brighter in unpolarised light, as expected from enhanced forward scattering. We therefore confirm that this side is inclined towards the Earth. This reconciles the polarised and unpolarised view of the disc, without the need for a marginally optically thick ring, as proposed by  \citet{Perrin2015}.

The SPF of the dust shows two regimes. The first one up to $45^\circ$ is dominated by forward scattering, and the second one above $45^\circ$ displays a smooth increasing SPF with scattering angle dominated by backward scattering. It cannot be reproduced by simple Mie or DHS models, but points towards large aggregates. 

This peak of forward scattering is compatible with the diffraction peak by 20\micron{} particles. Mie or DHS grains of that size fail to reproduce the SED, but again aggregates could have the same opacity as smaller grains if the filling factor is also smaller.

We provide a reflectance spectrum of the dust with a resolution of 50 between 0.95\micron{} and 1.35\micron, confirming the red spectra of the dust already observed with HST. We do not see any additional absorption features within our error bars.

We do not detect any companions orbiting within or outside the disc but considerably improve the detection limits set on this system by previous high-contrast near-infrared observations at very short separation, for example, from 32 to $5M_\text{Jup}$ for instance at 0.2\arcsec. Combining our detection limits with predictions from dynamical models enable us to exclude the presence of planets more massive than 2$M_\text{Jup}$ (respectively 3.6$M_\text{Jup}$) orbiting interior to the ring along the semi-major axis of the disc (respectively semi-minor axis).

After considering two plausible scenarios as the origin for the gas, either primordial or secondary, we set an upper limit on the gas-to-dust ratio of 0.2. This can rule out the presence of gas as the origin of the eccentricity and steepness of the ring. 

We conclude by emphasising that both an observational and theoretical effort is needed to fully reconcile both the spectral reflectance and the thermal emission. The presence of large aggregates showing different behaviours whether observed in thermal or scattered light regime, and in polarised or unpolarised light is the solution we propose here and that is supported by these new high-contrast and high-angular resolution observations. 

\begin{acknowledgements}
J.M. acknowledges financial support from the ESO fellowship programme. AML acknowledges the support from the ANR blanche GIPSE (ANR-14-CE33-0018) and the Labex OSUG. LM acknowledges support by STFC and ESO through graduate studentships and by the European Union through ERC grant number 279973. We would like to thank ESO staff and technical operators at the Paranal Observatory. We thank M. Meyer and D. Rouan for their valueable suggestions and comments during the review by the SPHERE internal board. We thank P. Delorme and E. Lagadec (SPHERE Data Center) for their work during the data reduction process. We thank V. Faramaz for the discussion on the eccentricity of the disc. SPHERE is an instrument designed and built by a consortium consisting of IPAG (Grenoble, France), MPIA (Heidelberg, Germany), LAM (Marseille, France), LESIA (Paris, France), Laboratoire Lagrange (Nice, France), INAF - Osservatorio di Padova (Italy), Observatoire de Geneve (Switzerland), ETH Zurich (Switzerland), NOVA (Netherlands), ONERA (France) and ASTRON (Netherlands) in collaboration with ESO. SPHERE was funded by ESO, with additional contributions from CNRS (France), MPIA (Germany), INAF (Italy), FINES (Switzerland) and NOVA (Netherlands). SPHERE also received funding from the European Commission Sixth and Seventh Framework Programmes as part of the Optical Infrared Coordination Network for Astronomy (OPTICON) under grant number RII3-Ct-2004-001566 for FP6 (2004-2008), grant number 226604 for FP7 (2009-2012) and grant number 312430 for FP7 (2013-2016).
\end{acknowledgements}

%-------------------------------------------------------------------
\bibliography{biblio_HR4796}    

\Online
\begin{appendix} 

\section{MCMC fit of the elllipse}
\label{App_MCMC}

\begin{figure*}
        \centering
        \includegraphics[width=\hsize]{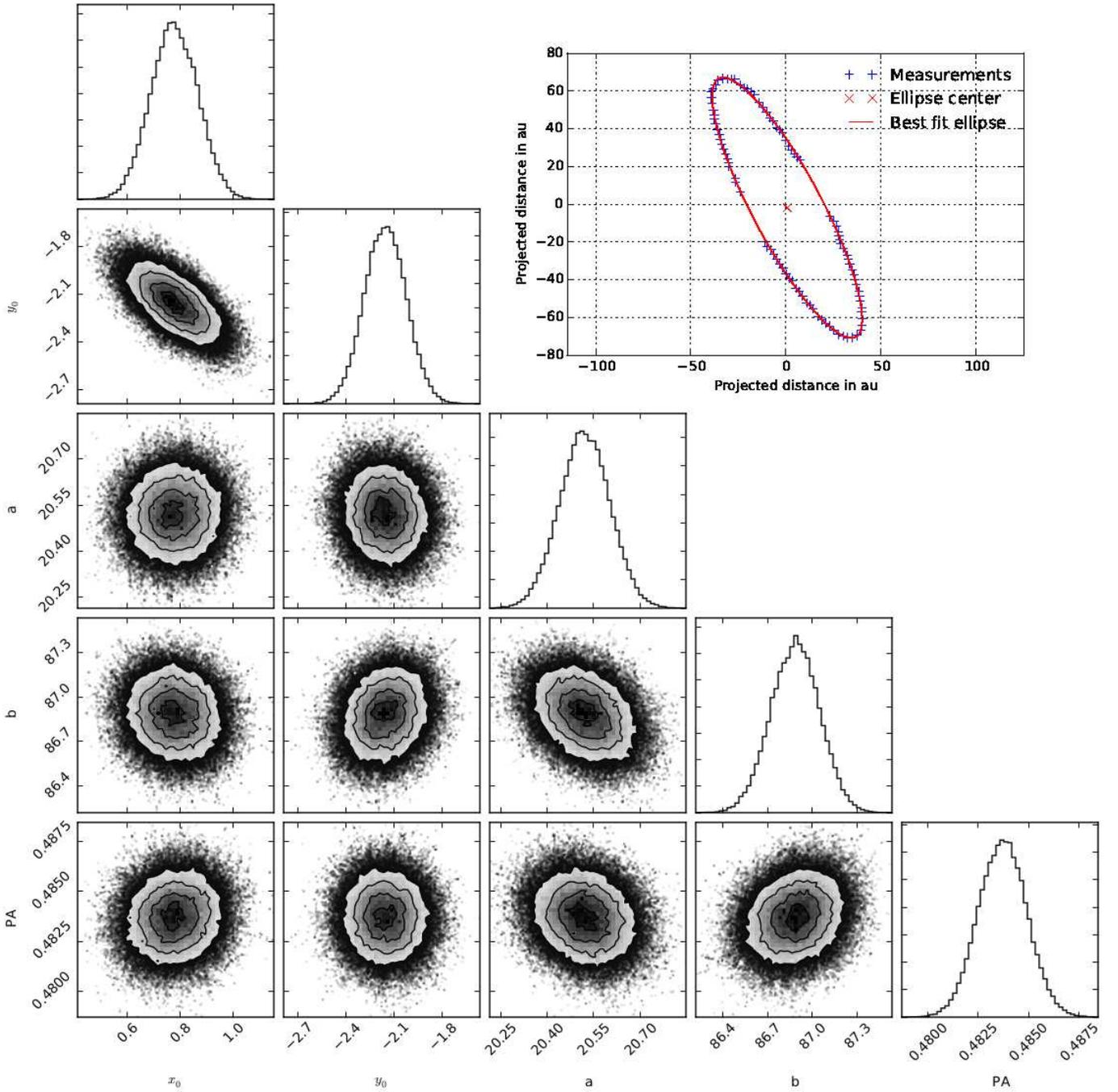}
        \caption{Full probability density distribution of the projected parameters of the ellipse fitted to the IRDIS image in H band. The image in the top right hand corner shows the data points and the corresponding best fit ellipse.}
        \label{fig_triangle_irdis_H}
\end{figure*}

\begin{figure*}
        \centering
        \includegraphics[width=\hsize]{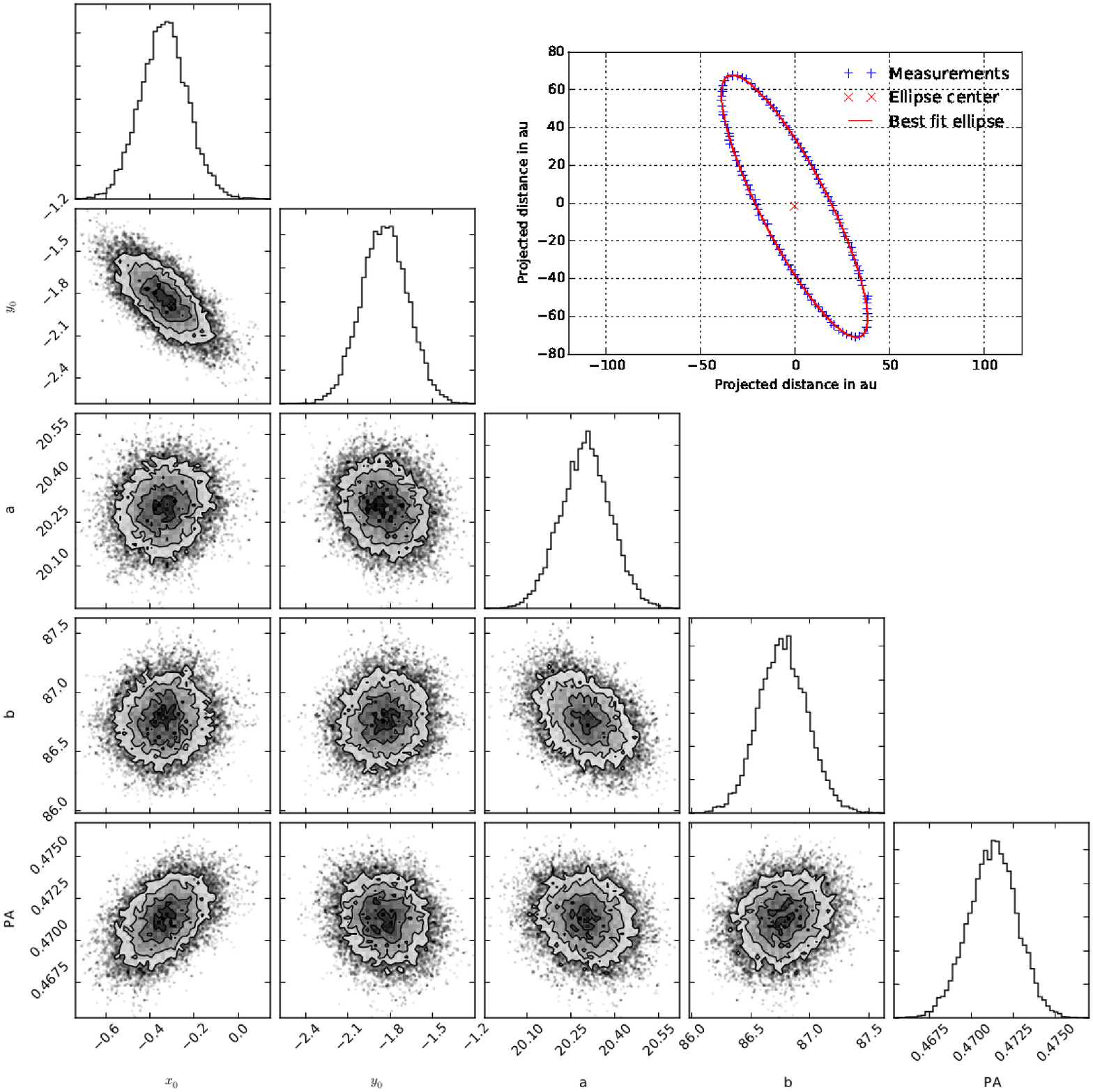}
        \caption{Full probability density distribution of the projected parameters of the ellipse fitted to the IRDIS image in the H2 band. The image in the top right hand corner shows the data points and the corresponding best fit ellipse.}
        \label{fig_triangle_irdis_H2}
\end{figure*}

\begin{figure*}
        \centering
        \includegraphics[width=\hsize]{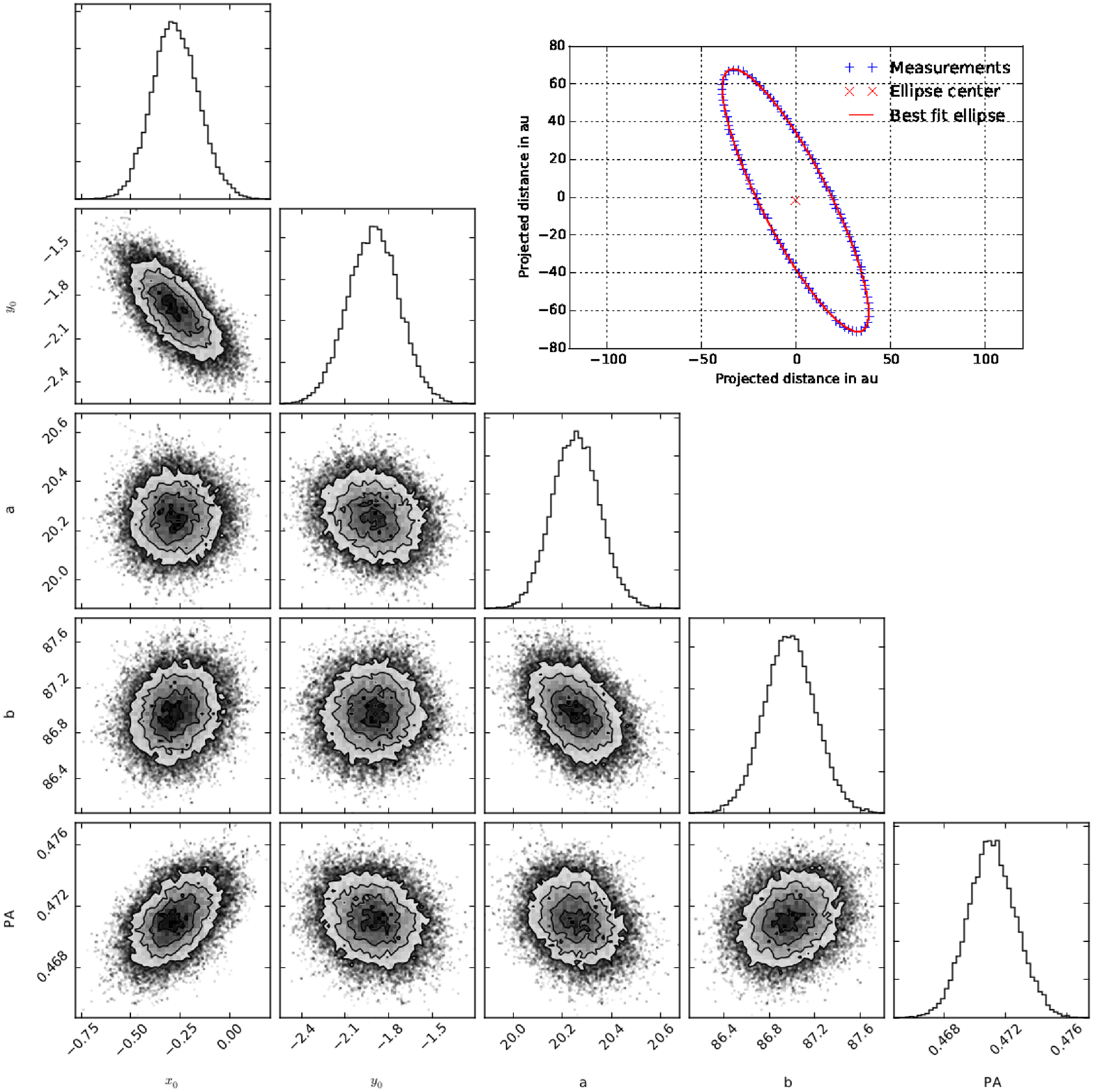}
        \caption{Full probability density distribution of the projected parameters of the ellipse fitted to the IRDIS image in the H3 band. The image in the top right hand corner shows the data points and the corresponding best fit ellipse.}
        \label{fig_triangle_irdis_H3}
\end{figure*}

\begin{figure*}
        \centering
        \includegraphics[width=\hsize]{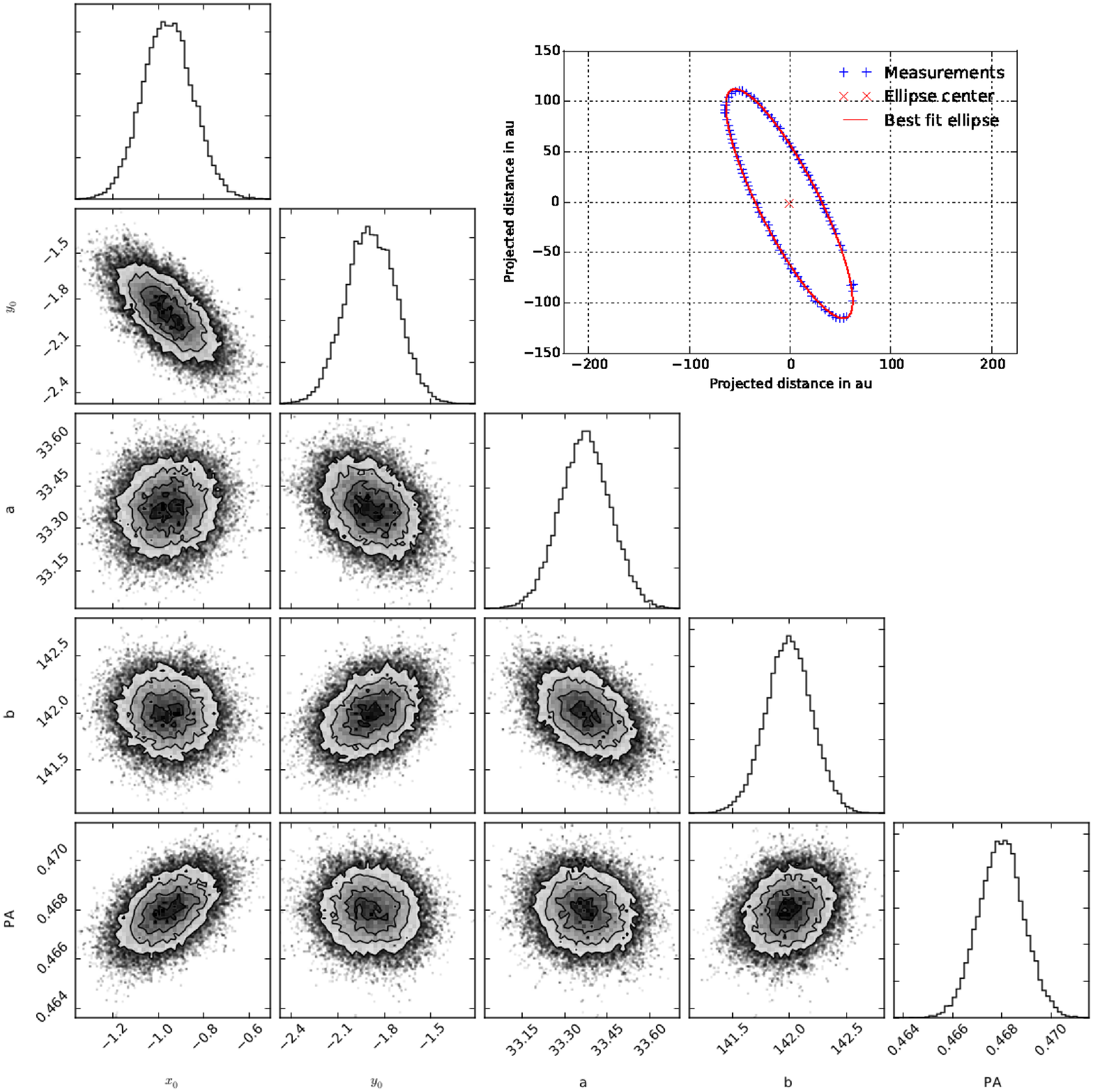}
        \caption{Full probability density distribution of the projected parameters of the ellipse fitted to the IFS image (collapsed spectral channels from Y to J). The image in the top right hand corner shows the data points and the corresponding best fit ellipse.}
        \label{fig_triangle_IFS}
\end{figure*}

An ellipse is entirely characterised by the five following parameters : the centre coordinates ($x_0$,$y_0$), the semi-major and semi-minor axis $a$ and $b$ and the position angle of the semi-major axis $PA$. \citet{Ray2008} proposed a geometric approach to characterize the misfit between an ellipse parametrised by the model vectior $\mathbf{u}=(x_0,y_0,a,b,PA)$ and a set of data points. We adopted their definition: if $F_i(\mathbf{u})$ is the distance between the $i^{th}$  data point and its projection on the ellipse as defined in their Fig. 3, the misfit is the sum of $F_i^2(\mathbf{u})$. Finding the minimum misfit is a nonlinear least square problem that we chose to solve with a Markov chain Monte Carlo technique (MCMC). We implemented the affine-invariant ensemble sampler called emcee \citep{Foreman-Mackey2013}. We assumed uniform priors for each ellipse parameter. The posterior probability density function is shown in Figs. \ref{fig_triangle_irdis_H}, \ref{fig_triangle_irdis_H2}, \ref{fig_triangle_irdis_H3} and \ref{fig_triangle_IFS}, for the H, H2, H3 and IFS image, as well as the data point and best model ellipse (inset image in the top right hand corner).

\section{Sanity check of the derivation of the ring morphological parameters}
\label{App_sanity_check}

As our goal in this case is purely to derive the geometrical properties of the ring, we used an isotropic model with a dust density distribution parametrised by the following expression in cylindrical coordinates $(r,\theta,z)$ \citep{Augereau1999}:
\begin{equation}
\label{eq_dust_density}
\rho(r,\theta,z) = \rho_0 \times \left( \frac{2}{\left( \frac{r}{R(\theta)} \right)^{-2a_{in}} + \left( \frac{r}{R(\theta)} \right)^{-2a_{out}} }\right)^{1/2} \times e^{\left[ -\left( \frac{z}{H(r) }\right)^\gamma \right]}
,\end{equation}

where the reference radius $R$ is an ellipse parametrised in polar coordinates by the expression
\begin{equation}
\label{eq_eccentric_ref_radius}
R(\theta) = \frac{a(1-e^2)}{1+e \cos{\theta}}
.\end{equation}

In case of a non eccentric orbit, $e=0$ and $R$ is independent of $\theta$. The scale height of the disc ($H$), is written as
\begin{equation}
\label{eq_vertical_height}
H(r)=\xi_0 \left( \frac{r}{a(1-e^2)}\right)^\beta
.\end{equation}
To be consistent with previous models, for example, \citet{Milli2015}, we assumed a linear flaring $\beta=1$ and a Gaussian vertical profile $\gamma=2$ of reference height $\xi_0=1$ au. We are therefore left with five free parameters ($a$,$e$,$i$,$\omega$,$\Omega$), $\rho_0$ being directly scaled to the image in the assumption of an optically thin disc.
We used again the emcee MCMC sampler to find the best solution that maximizes the likelihood function defined as $-0.5e^{-\chi^2}$, where $\chi^2$ is the chi squared between the image and the model using a mask that shapes as an elliptical annulus adapted to the ring of HR\,4796. The result is shown in Table \ref{tab_ring_morphological_sanity_check}  in the rows corresponding to deprojected image. They agree very well with the parameters derived from the fit of the discrete sampling of the ring described earlier. The uncertainty is about six times higher, this is likely related to the fact that the model was assumed to be isotropic to limit the parameter space\footnote{The MCMC uses 500 steps and 100 walkers, it requires about three days on a standard CPU of a MacBook Pro.} which is clearly not the case. 

\begin{table*}
\caption{Deprojected ring parameters as measured using an alternative technique. The error is given at a $1\sigma$ level and contains only the statistical error from the fit and no systematic error from the true north or star registration.}
\label{tab_ring_morphological_sanity_check}
\centering
\begin{tabular}{c c |c c c c c c}
\hline 
\multicolumn{2}{c}{Type of fit} & Parameter &  IRDIS H & IRDIS H2 & IRDIS H3 & IFS \\
\hline
\hline
\parbox[t]{2mm}{\multirow{5}{*}{\rotatebox[origin=c]{90}{Deprojected}}} & \parbox[t]{2mm}{\multirow{5}{*}{\rotatebox[origin=c]{90}{image\tablefootmark{a}}}} & $a$(mas) & $ 1064 \substack{+   11 \\ -   11}$  & $ 1065 \substack{+   30 \\ -   30}$ & $ 1067 \substack{+   23 \\ -   25}$ &  $ 1072 \substack{+   54 \\ -   65}$ \\
                                                     & & $e$ &   $0.066 \substack{+0.030 \\ -0.029}$ & $0.039 \substack{+0.049 \\ -0.028}$ &  $0.031 \substack{+0.039 \\ -0.022}$ &  $0.077 \substack{+0.075 \\ -0.055}$ \\
                                                     & & $i$($^\circ$)  &  $76.43 \substack{+ 0.70 \\ - 0.68}$ &$76.52 \substack{+ 1.35 \\ - 1.39}$ & $76.62 \substack{+ 0.95 \\ - 1.02}$ & $75.69 \substack{+ 3.30 \\ - 4.79}$  \\
                                                     & & $\omega$($^\circ$) &  $-73.52\substack{+15.31 \\ - 9.25}$ & $-53.24\substack{+119.62 \\ -42.83}$ & $-55.32\substack{+111.28 \\ -40.69}$ & $-22.27\substack{+124.34 \\ -93.15}$  \\
                                                     & & $\Omega$($^\circ$) & $27.78 \substack{+ 0.53 \\ - 0.50}$ & $27.01 \substack{+ 1.14 \\ - 1.21}$ & $26.91 \substack{+ 0.96 \\ - 0.93}$     & $26.82 \substack{+ 4.62 \\ - 4.90}$ \\
\hline
\end{tabular}
\end{table*}

 \end{appendix}

\end{document}